\documentclass[a4paper,twoside]{article}
\newcommand{\affil}[1]{$^{\rm #1}$}
\usepackage[authoryear]{natbib}
\bibpunct{(}{)}{;}{a}{}{,}
\usepackage{graphicx}
\usepackage{upgreek}
\usepackage{amssymb}
\date{} %Please leave the date blank
\newcommand{\ion}[2]{#1$\,$\textsc{#2}}

\newcommand{\fdeg}{\mbox{\ensuremath{.\!^{\circ}}}}
\title{\large\bf\flushleft Basic Testing of the {\sc Duchamp} Source Finder}
\author{\parbox{\textwidth}{\flushleft
\vspace{-0.5cm}
{\it T.~Westmeier\affil{A,C}, A.~Popping\affil{A}, and P.~Serra\affil{B}}\\
\vspace{0.4cm}
{\small \affil{A}\,International Centre for Radio Astronomy Research, M468, The University of Western Australia, 35~Stirling Highway, Crawley WA 6009, Australia}\\
{\small \affil{B}\,ASTRON, Postbus~2, 7990~AA~Dwingeloo, the Netherlands}\\
{\small \affil{C}\,Email: tobias.westmeier@uwa.edu.au}}}
\begin{document}
\twocolumn[
\begin{changemargin}{.8cm}{.5cm}
\begin{minipage}{.9\textwidth}
\vspace{-1cm}
\maketitle
%
%
%%%%%%%%%%%%%     ABSTRACT    %%%%%%%%%%%%%
%Abstract of no more than 200 words here.
\small{\bf Abstract:} This paper presents and discusses the results of basic source finding tests in three dimensions (using spectroscopic data cubes) with \textsc{Duchamp}, the standard source finder for the Australian SKA Pathfinder. For this purpose, we generated different sets of unresolved and extended \ion{H}{i} model sources. These models were then fed into \textsc{Duchamp}, using a range of different parameters and methods provided by the software. The main aim of the tests was to study the performance of \textsc{Duchamp} on sources with different parameters and morphologies and assess the accuracy of \textsc{Duchamp}'s source parametrisation. Overall, we find \textsc{Duchamp} to be a powerful source finder capable of reliably detecting sources down to low signal-to-noise ratios and accurately measuring their position and velocity. In the presence of noise in the data, \textsc{Duchamp}'s measurements of basic source parameters, such as spectral line width and integrated flux, are affected by systematic errors. These errors are a consequence of the effect of noise on the specific algorithms used by \textsc{Duchamp} for measuring source parameters in combination with the fact that the software only takes into account pixels above a given flux threshold and hence misses part of the flux. In scientific applications of \textsc{Duchamp} these systematic errors would have to be corrected for. Alternatively, \textsc{Duchamp} could be used as a source finder only, and source parametrisation could be done in a second step using more sophisticated parametrisation algorithms.

%%%%%%%%%%%%%     KEYWORDS    %%%%%%%%%%%%%
\medskip{\bf Keywords:} methods: data analysis
% Please write all keywords in lower case. PASA uses the
% standard list of subject headings adopted by The Astrophysical Journal
% and available from http://www.journals.uchicago.edu/ApJ/keywords_text.html.
% Keywords are separated by em-dashes, i.e. ---

%%%%%%%%DO NOT EDIT%%%%%%%%%%%%
\medskip
\medskip
\end{minipage}
\end{changemargin}
]
\small
%%%%%%%%EDIT FROM HERE%%%%%%%%%%%%

  \section{Introduction}
  
  With the advent of the \textit{Square Kilometre Array} (SKA; \citealt{dewdney2009}) and its precursors and pathfinders, including the \textit{Australian SKA Pathfinder} (ASKAP; \citealt{deboer2009}), the \textit{Karoo Array Telescope} (Meer\-KAT; \citealt{jonas2009}), and the \textit{Aperture Tile In Focus} (APERTIF; \citealt{oosterloo2009}), the prospect of deep radio continuum and \ion{H}{i} surveys of large areas on the sky demands for new strategies in the areas of data reduction and analysis, given the sheer volume of the expected data streams, in particular for spectroscopic surveys.
  
  Of particular importance is the automatic and accurate identification and parametrisation of sources with high completeness and reliability. Due to the large data volumes to be searched, source finding algorithms must be fully automated, and the once common practice of source finding `by eye' will no longer be feasible. Moreover, accurate source para\-metrisation algorithms need to be developed to generate reliable source catalogues free of systematic errors, as otherwise the integrity of scientific results based on the survey data could be compromised.
  
  In this paper we will take a closer look at the \textsc{Duchamp} source finder\footnote{\textsc{Duchamp} website: http://www.atnf.csiro.au\slash{}people\slash{}Matthew.Whiting\slash{}Duchamp/} \citep{whiting2011a,whiting2012}. \textsc{Du\-champ} has been developed by Matthew Whiting at CSIRO as a general-purpose source finder for three-dimensional data cubes as well as two-di\-mensional images and will serve as the default source finder in the processing of data from the ASKAP survey science projects. The software identifies sources by searching for regions of emission above a specified flux threshold. To improve its performance, \textsc{Duchamp} offers several methods of preconditioning and filtering of the input data, including spatial and spectral smoothing as well as reconstruction of the entire image or data cube with the help of wavelets. In addition to source finding, \textsc{Duchamp} provides the user with basic source parametrisation, including the measurement of position, size, radial velocity, line width, and integrated flux of a source. More information about the capabilities of the software is available from the \textsc{Duchamp} User Guide \citep{whiting2011b}. A brief overview of \textsc{Du\-champ}'s basic functionality is provided in Section~\ref{sect_duchamp}.
  
  \begin{table}[ht]
    \begin{center}
      \caption{Summary of the parameters used to generate the visibility data set and noise image for the point source models.}
      \label{tab_model}
      \begin{tabular}{lrl}
        \hline
        Parameter (visibility) &         Value & Unit           \\
        \hline
        Number of antennas     &           $36$ &                \\
        System temperature     &           $50$ & $\mathrm{K}$   \\
        Declination            &  $-45^{\circ}$ &                \\
        Total integration time &            $8$ & $\mathrm{h}$   \\
        Hour angle range       &        $\pm 4$ & $\mathrm{h}$   \\
        Cycle time             &            $5$ & $\mathrm{s}$   \\
        Stokes parameters      &              I &                \\
        Number of channels     &           $31$ &                \\
        Frequency              &         $1.42$ & $\mathrm{GHz}$ \\
        Channel width          &        $18.31$ & $\mathrm{kHz}$ \\
                               &         $3.86$ & $\mathrm{km \, s}^{-1}$ \\
        \hline
        Parameter (image)      &          Value & Unit                 \\
        \hline
        Final image size       & $31 \times 31$ & $\mathrm{px}$        \\
        Field diameter         &            $5$ & $\mathrm{arcmin}$    \\
        Pixel size             &           $10$ & $\mathrm{arcsec}$    \\
        Robustness             &            $0$ &                      \\
        Gaussian $uv$ taper    &         $7.28$ & $\mathrm{k} \lambda$ \\
                               &         $1.54$ & $\mathrm{km}$        \\
        \textsc{rms} noise     &         $1.95$ & $\mathrm{mJy}$       \\
        Synthesised beam       &                &                      \\
        \quad major axis       &         $27.1$ & $\mathrm{arcsec}$    \\
        \quad minor axis       &         $26.7$ & $\mathrm{arcsec}$    \\
        \quad position angle   &   $87\fdeg{}9$ &                      \\
        \hline
      \end{tabular}
    \end{center}
  \end{table}
  
  So far, \textsc{Duchamp}'s source finding and parametrisation capabilities have never been systematically tested on a large set of artificial sources with well-defined parameters. The aim of this paper is to bridge this gap by thoroughly testing the performance of \textsc{Duchamp} on sets of artificial point sources and galaxy models as well as a data set containing real galaxies and telescope noise. The tests were originally motivated by the need to identify suitable source finding algorithms for the \textit{Widefield ASKAP L-band Legacy All-sky Blind Survey} (WALLABY; \citealt{koribalski2009}),\footnote{Principal investigators: B\"{a}rbel Koribalski and Lister Staveley-Smith; public website: http://www.atnf.csiro.au\slash{}research\slash{}WALLABY/} one of the large, extragalactic ASKAP survey science projects currently in pre\-paration \citep{westmeier2010}. Hence, the tests presented here will focus on the detection of compact and extended \ion{H}{i}~sources, in particular galaxies, in three-dimensional data cubes with ASKAP characteristics.
  
  However, we believe that the results and conclusions presented in this paper will be of interest not only to those involved in SKA precursor science, but to a larger community of astronomers interested in the automatic detection and parametrisation of sources in their data sets, regardless of the wavelength range involved. For a comparison of \textsc{Duchamp}'s performance with that of other source finding algorithms we refer to the paper by \citet{popping2011} in this issue.
  
  This paper is organised as follows: In Section~\ref{sect_surveys} we summarise the source finding strategies of other large \ion{H}{i}~surveys in the past, followed by a brief overview of the \textsc{Duchamp} source finder in Section~\ref{sect_duchamp}.  In Section~\ref{sect_pointsources} we present the outcome of our test of \textsc{Duchamp} on point source models with simple Gaussian line profiles. Section~\ref{sect_galaxies} describes our testing of \textsc{Duchamp} on models of disc galaxies with varying physical parameters. In Section~\ref{sect_whisp} we apply \textsc{Duchamp} to a data cube containing real galaxies and genuine noise extracted from radio observations. A discussion of our results is presented in Section~\ref{sect_discussion}. Finally, Section~\ref{sect_summary} summarises our main results and conclusions.
  
  \section{Source Finding in Previous Surveys}
  \label{sect_surveys}
  
  Some of the previous, large \ion{H}{i}~surveys, including the \ion{H}{i}~Parkes All-Sky Survey (HIPASS; \citealt{barnes2001}), the \ion{H}{i}~Jodrell All-Sky Survey (HIJASS; \citealt{lang2003}), and the Arecibo Legacy Fast ALFA survey (ALFALFA; \citealt{giovanelli2005}), already had to deal with the issue of (semi-)automatic source detection.
  
  In the case of HIPASS, two different source finders were used and the results combined to maximise completeness \citep{meyer2004}. The first algorithm, \textsc{multifind} \citep{kilborn2001}, used a simple $4 \sigma$ flux threshold combined with smoothing of the data cube on different scales. The second algorithm, \textsc{tophat}, detected sources in the spectral domain by convolving each spectrum in the data cube with a top-hat function of varying width. Neither of the two algorithms alone managed to detect more than 90\% of the final, combined source list. The two algorithms combined produced about 140,000 unique detections, all of which were inspected by eye to remove potential false detections. The final HIPASS catalogue included 4315 sources \citep{meyer2004}, resulting in an overall reliability of the automatic source finding algorithms of only about 3\%.
  
  In the case of HIJASS, again two different methods were used and the results combined to improve completeness \citep{lang2003}. The first method simply involved searching the cubes by eye to extract potential sources. The second algorithm, \textsc{polyfind}, first searched for signals above a given threshold in a smoothed version of the data cube and then ran a series of matched filters over the detected signals to decide whether a signal was likely genuine or false. As in the case of HIPASS, the source list produced by the automatic source finding routine was inspected by eye to further reject potential false detections. The positions of uncertain detections were re-observed to either confirm of refute them.
  
  For the ALFALFA survey, a matched-filtering technique was applied to the data in the spectral domain \citep{saintonge2007}. The data were convolved with a set of template functions created by combining the first two symmetrical Hermite functions, $\Psi_{0}(x)$ and $\Psi_{2}(x)$. The resulting templates range from simple Gaussian profiles for narrow signals to double-peaked profiles for broader signals, covering the range of spectral shapes expected from \ion{H}{i}~observations of real galaxies. In tests on 1500~simulated galaxies, 100\% reliability and about 70\% completeness are achieved at an integrated signal-to-noise ratio of $S/N \approx 6$, while the 90\% completeness level is exceeded at $S/N \gtrsim 9$.\footnote{In her calculation of $S/N$, \citet{saintonge2007} makes the implicit assumption that the sources are spatially unresolved.}
  
  \section{The {\sc Duchamp} Source Finder}
  \label{sect_duchamp}
  
  \textsc{Duchamp} has been implemented as a general-purpose source finder for three-dimensional spectral-line data cubes with two spatial axes and one frequency (or velocity) axis, although the software can also operate on one- and two-dimensional data sets \citep{whiting2011b}. \textsc{Duchamp} finds sources by applying a simple flux threshold to the data cube, specified by the \texttt{threshold} or \texttt{snrCut} keyword, and searching for pixels above that threshold. In a second step, the software attempts to merge detections into sources under specific circumstances that can be controlled by the user. One option is to simply merge adjacent pixels (\texttt{flagAdjacent} keyword). Alternatively, a maximum spatial and spectral separation can be specified for the merging of detected pixels into sources (\texttt{threshSpatial} and \texttt{threshVelocity} keywords, respectively). Once detected, sources can be ``grown'' to a flux level below the actual detection threshold, using the \texttt{flagGrowth} and \texttt{growthThreshold}/\texttt{growthCut} keywords.
  
  Basic removal of false detections is achieved by requiring that sources comprise a minimum number of contiguous spatial pixels and spectral channels, using the \texttt{minPix} and \texttt{minChannels} keywords, respectively. To improve the reliability of the source finding even further, \textsc{Duchamp} offers a powerful method of reconstructing the entire input data cube with the help of wavelets. Source finding is then performed on the reconstructed cube instead of the original input cube. Reconstruction can either be carried out in all three dimensions of the cube, or in the spatial (two-dimensional) or spectral (one-di\-mensional) domain only.
  
  \textsc{Duchamp} uses the so-called `\`{a} trous' wavelet reconstruction method \citep{starck2002}. First, the input data set is convolved with a specific wavelet filter function (three different functions are offered to the user by \textsc{Duchamp}). The difference between the convolved data set and the original data set is then added to the reconstructed cube. Next, the scale of the filter function is doubled and the procedure repeated, using the convolved array as the new input data set. Once the user-specified maximum filter scale is reached, the final convolved data set is added to the reconstructed cube, and source finding on the reconstructed data set commences.
  
  The `\`{a} trous' wavelet reconstruction of the data cube offers a powerful method of enhancing \textsc{Duchamp}'s source finding capabilities. First of all, the user can select the minimum (\texttt{scaleMin} keyword) and maximum (\texttt{scaleMax} keyword) filter scales to be used in the reconstruction, providing efficient suppression of small-scale and large-scale artefacts in the data, such as noise peaks, baseline ripples, or radio-frequency interference. Furthermore, the user can specify an additional threshold (\texttt{snrRecon} keyword) to be applied when adding wavelet components to the reconstructed data cube, thereby reducing even further the number of spurious signals in the data cube.
  
  In comparison to simple data thresholding, the `\`{a} trous' wavelet reconstruction method will greatly increase the completeness and reliability of \textsc{Duchamp}'s source finding procedure, and hence the method has been applied in all source finding tests presented in this paper.
  
  \begin{figure}[t]
    \begin{center}
      \includegraphics[width=\linewidth]{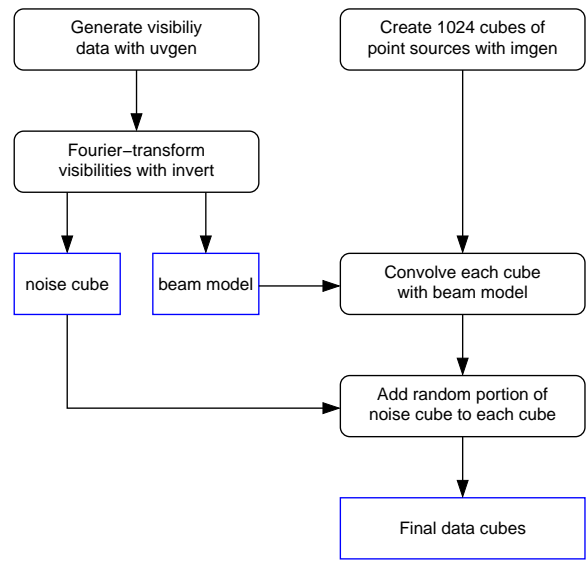}
      \caption{Outline of the procedure used to create the model data cubes of point sources with \textsc{Miriad}.}
      \label{fig_procedure}
    \end{center}
  \end{figure}
  
  \section{Point Sources with Gaussian Spectral Profiles}
  \label{sect_pointsources}
  
  For our first test of \textsc{Duchamp} we generated models of 1024~point sources with simple Gaussian spectral line profiles. This will allow us to assess the fundamental performance of \textsc{Duchamp} under ideal conditions and to investigate the accuracy of the software's source parametrisation algorithms. Point sources with Gaussian profiles are ideal for this test because---as a consequence of their simple morphology---their physical parameters can be exactly defined and calculated to serve as a benchmark for \textsc{Duchamp}'s parametrisation.
  
  In order to create the model data set, the \textsc{Miriad} \citep{sault1995} task \textsc{uvgen} was employed to generate visibility data of Gaussian noise at a frequency of $1.4~\mathrm{GHz}$ with ASKAP characteristics and parameters similar to those anticipated for the WALLABY survey. The model parameters are summarised in Table~\ref{tab_model}.
  
  \begin{figure*}[t]
    \begin{center}
      \includegraphics[width=0.9\linewidth]{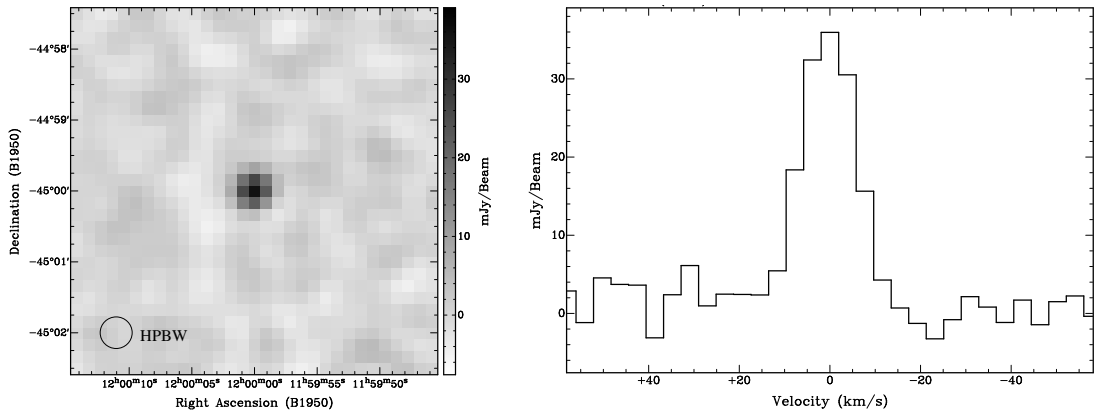}
      \caption{Example of a point source model generated for testing \textsc{Duchamp}. The left-hand panel shows a single-channel map of the data cube (at the systemic velocity of the source), and the right-hand panel depicts the spectrum at the source position. The circle in the map illustrates the half-power beam width.}
      \label{fig_maps}
    \end{center}
  \end{figure*}
  
  The visibility data were Fourier-transformed using \textsc{Miriad}'s task \textsc{invert} to generate a noise image of $600 \times 600$~pixels and 31~spectral channels with characteristics similar to WALLABY (again, see Table~\ref{tab_model} for details). The \textsc{rms} noise level in this image is $\sigma = 1.95~\mathrm{mJy}$ which is only slightly higher than the $1.6~\mathrm{mJy}$ expected for WALLABY.
  
  In order to generate images of point sources, the \textsc{Miriad} task \textsc{imgen} was used to create 1024~data cubes each of which has a size of $31 \times 31$~pixels and 31~spectral channels and contains a single point source with Gaussian spectral line profile in the centre. Each source was randomly assigned a peak flux in the range of $1$ to $20 \sigma$, resulting in an average of about 54~sources per $1 \sigma$ interval. Spectral line widths (FWHM) range from $0.1$ to $10$~spectral channels, equivalent to approximately $0.4$ to $38.6~\mathrm{km \, s}^{-1}$, resulting in a density of about 27~sources per $1~\mathrm{km \, s}^{-1}$ line width interval. While in reality sources with \ion{H}{i}~line widths of as small as $0.4~\mathrm{km \, s}^{-1}$ will not exist, the reason for including such narrow lines in our test is to study the performance of \textsc{Duchamp} on sources that are spectrally unresolved, irrespective of the absolute line width.
  
  Each of the 1024~cubes was convolved with the beam model produced by \textsc{invert}. Next, a random portion of $31 \times 31$~pixels of the original noise cube was selected and added to each convolved image to create the final images used for testing \textsc{Duchamp}. To facilitate correct integrated flux measurements, we added information on the synthesised beam to the image header. The entire procedure is outlined in Figure~\ref{fig_procedure}. An example image and spectrum of one of the point source models is shown in Figure~\ref{fig_maps}.
  
  \begin{table}[b]
    \begin{center}
      \caption{Relevant input parameters for the different test runs of \textsc{Duchamp} in order to find the optimal set of control parameters (see Figure~\ref{fig_tests}). The best-performing parameter set (run~5) was then used for the analysis presented in this paper.}
      \label{tab_duchamp_tests}
      \begin{tabular}{ccc}
        \hline
        Run & threshold & scaleMin \\
        \hline
        1   &     $1.5$ &        1 \\
        2   &     $2.0$ &        1 \\
        3   &     $2.5$ &        1 \\
        4   &     $3.0$ &        1 \\
        5   &     $1.5$ &        2 \\
        6   &     $2.0$ &        2 \\
        7   &     $1.0$ &        3 \\
        8   &     $0.5$ &        3 \\
        \hline
      \end{tabular}
    \end{center}
  \end{table}
  
  \begin{table}[b]
    \begin{center}
      \caption{\textsc{Duchamp} input parameters \citep{whiting2011b} explicitly set in the input parameter file for point source models. The default values of \textsc{Duchamp} were used for all other parameters.}
      \label{tab_input}
      \begin{tabular}{lrl}
        \hline
        Parameter    &     Value & Comment      \\
        \hline
        threshold    & 0.0029265 & 1.5 $\times$ \textsc{rms} \\
        minPix       &         5 &              \\
        minChannels  &         3 &              \\
        flagAdjacent &      true &              \\
        flagATrous   &      true & Wavelet reconstr. \\
        reconDim     &         3 & in 3 dimensions \\
        snrRecon     &         3 &              \\
        scaleMin     &         2 &              \\
        \hline
      \end{tabular}
    \end{center}
  \end{table}
  
  \subsection{Running {\sc Duchamp}}
  
\begin{figure}[t]
    \begin{center}
      \includegraphics[width=\linewidth]{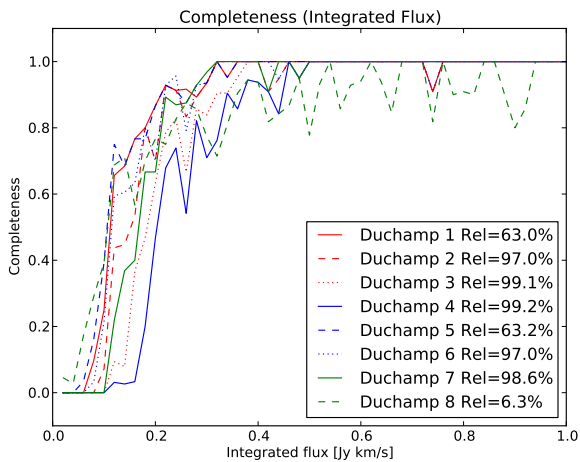}
      \caption{Completeness as a function of integrated flux for different tests of \textsc{Duchamp} with varying control parameters. The parameters employed in the different runs are listed in Table~\ref{tab_duchamp_tests}. The overall reliability for each run is listed in the legend.}
      \label{fig_tests}
    \end{center}
  \end{figure}

  Next, we ran \textsc{Duchamp} (version~1.1.8) on the data cubes. In order to find out which combination of control parameters provided the best performance in terms of completeness and reliability, we first ran \textsc{Duchamp} several times with different flux thresholds and minimum wavelet scales to test the performance of each set of parameters. An overview of the different completeness and reliability levels achieved in these runs as a function of integrated flux of the source is shown in Figure~\ref{fig_tests}.
  
  We then selected the best set of control parameters for the analysis presented in this section. In this best-performing run (number~5 in Figure~\ref{fig_tests} and Table~\ref{tab_duchamp_tests}) we used a $1.5 \sigma$~flux threshold equivalent to $2.9~\mathrm{mJy}$. In addition, we made use of \textsc{Duchamp}'s `{\`a} trous' wavelet reconstruction. We employed a full three-dimen\-sional wavelet reconstruction with a minimum wavelet scale of~$2$ (i.e.~the smallest scales were excluded to suppress noise in the reconstructed cube) and a flux threshold of $3 \sigma$ for wavelet components to be included in the reconstructed cube. In addition, we required sources to cover a minimum of 5~contiguous pixels in the image domain and 3~contiguous spectral channels above the detection threshold to be included in the final source catalogue. This will further reduce the number of spurious detections. The \textsc{Duchamp} input parameters explicitly set in the parameter file are listed in Table~\ref{tab_input}.
  
  The 1024~output parameter files generated by \textsc{Du\-champ} were concatenated, and those source entries whose positions were within $\pm 1$~pixel of the nominal source position were considered as genuine detections and selected for further processing and analysis. The results of this analysis will be presented and discussed in the following sections.
  
  For a number of reasons it is not possible to specify the typical time it takes for \textsc{Duchamp} to process a certain amount of data. Firstly, the performance of \textsc{Duchamp} strongly depends on the exact choice of input parameters, including detection threshold, wavelet reconstruction choices, or settings related to merging and discarding of initial detections. Three-dimensional wavelet reconstruction of the input data cube, for example, is particularly computationally expensive. Secondly, the running time of \textsc{Duchamp} on a particular data cube will depend on a large number of details, including the number of sources in the cube, their size and morphology, and in principle even the number density of sources in the cube. Thirdly, recent updates of the software have resulted in a significant improvement of \textsc{Duchamp}'s performance, in particular compared to version~1.1.8 used for part of the testing presented in this paper.
  
  To get a basic idea of the impact of the aforementioned parameters on the running time of \textsc{Duchamp} we performed a few simple tests on a standard laptop computer with a state-of-the-art, dual-core $2.3~\mathrm{GHz}$ CPU (only one core at a time was actually engaged) and $4~\mathrm{GB}$ of physical memory.  We ran \textsc{Duchamp} several times with different parameters on our artificial noise data cube of $600 \times 600$ spatial pixels and $31$~spectral channels without any sources in it. Using a $5 \sigma$ detection threshold, \textsc{Duchamp} takes about $0.64~\mathrm{s}$ of CPU time to run, producing no detections. When performing a one-dimensional wavelet reconstruction in the spectral dimension prior to source finding, the running time increases by a factor of~$30$ to about $19~\mathrm{s}$. Full three-dimensional wavelet reconstruction is slower by a factor of~$120$, requiring about $77~\mathrm{s}$ to complete.
  
  As mentioned before, these numbers strongly depend on the number and nature of sources present in the cube. Decreasing the flux threshold to $3 \sigma$ without wavelet reconstruction results in $966$~detections (all of which are noise peaks) and increases the running time of \textsc{Duchamp} to about $3.5~\mathrm{s}$. Processing time will also increase with cube size. Doubling the cube size to $62$~spectral channels increases the running time by a factor of~$2$ without wavelet reconstruction, but by factors of $1.9$ and $2.6$ in the case of one-dimensional and three-dimensional reconstruction, respectively, indicating that an increase in cube size does not translate into a proportional increase in processing time when dealing with wavelet reconstruction.
  
  \begin{figure}[t!]
    \centering
    \includegraphics[width=\linewidth]{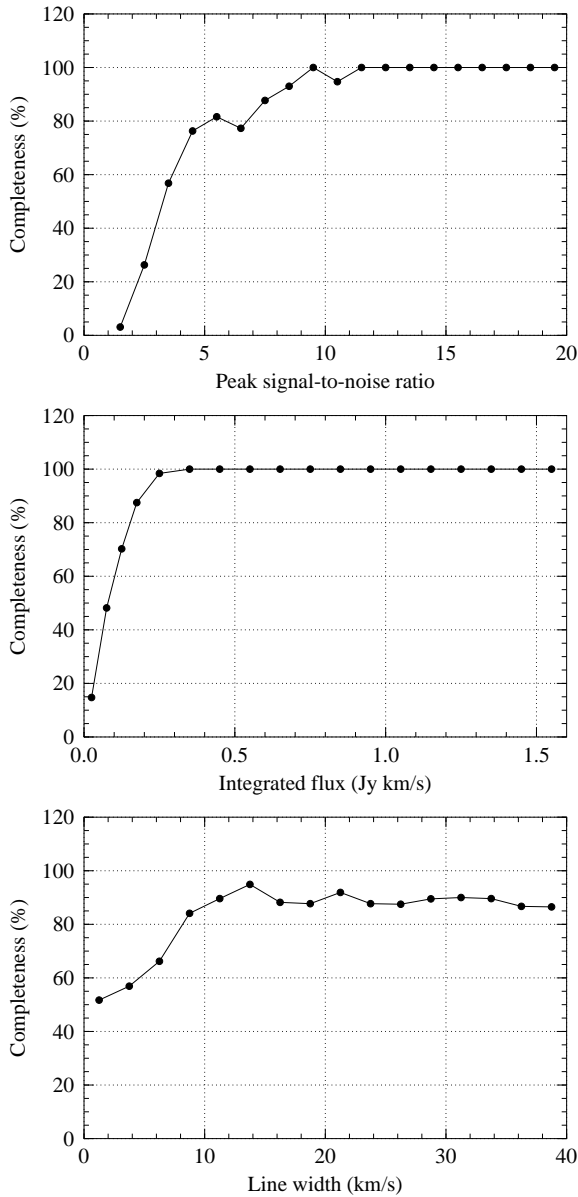}
    \caption{\textit{Top panel:} Completeness of the point source models as a function of true peak signal-to-noise ratio in bins of $1 \sigma$. \textit{Middle panel:} Same, but as a function of true integrated flux in bins of $0.1~\mathrm{Jy \, km \, s}^{-1}$. \textit{Bottom panel:} Same, but as a function of true line width (FWHM) in bins of $2.5~\mathrm{km \, s}^{-1}$.}
    \label{fig_completeness}
  \end{figure}
  
  \begin{figure}[t!]
    \centering
    \includegraphics[width=\linewidth]{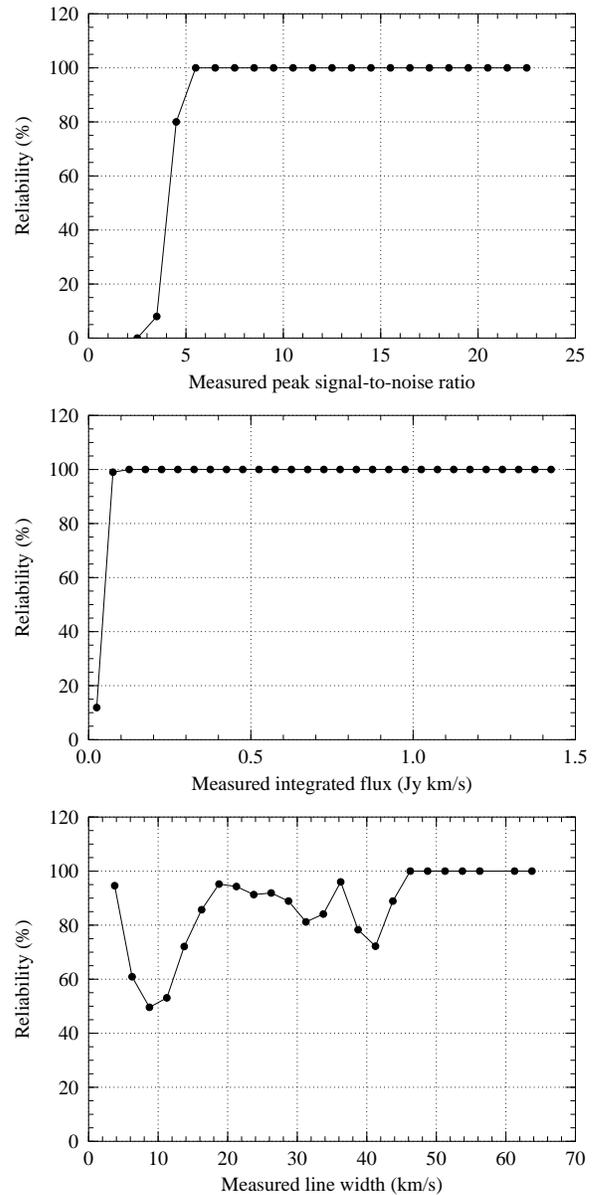}
    \caption{\textit{Top panel:} Reliability of the point source models as a function of measured peak signal-to-noise ratio in bins of $1 \sigma$. \textit{Middle panel:} Same, but as a function of measured integrated flux in bins of $0.05~\mathrm{Jy \, km \, s}^{-1}$. \textit{Bottom panel:} Same, but as a function of measured line width ($w_{50}$) in bins of $2.5~\mathrm{km \, s}^{-1}$.}
    \label{fig_reliability}
  \end{figure}
  
  In summary, the time \textsc{Duchamp} needs to process a data cube is a complicated function of not only the machine specifications (e.g.~CPU, memory, data transfer speed), but also the input parameters (e.g.~flux threshold, wavelet reconstruction) and properties of the data set concerned (e.g.~cube dimensions, number of sources). Hence, running times are almost impossible to predict and may have to be determined experimentally on a case-by-case basis. Instead of asking whether \textsc{Duchamp} is fast enough for a particular problem, the user would have to determine the optimal set of conditions that would allow processing of the data in a given period of time. An alternative option would be to separate the problem into multiple, parallel processes.
  
  \subsection{Results}
  
  \subsubsection{Completeness and Reliability}
  
  Two of the most important parameters in the characterisation of source finder performance are completeness and reliability. Completeness is defined as the number of genuine detections divided by the true number of sources present in the data. Completeness can either be calculated for the entire sample or more sensibly for a subset, e.g.\ for sources within a certain parameter range. Reliability is defined as the number of genuine detections divided by the total number of detections produced by the source finder. Reliability can only be calculated for the entire sample of sources and not for a subset of sources within a certain parameter range, because false detections do not possess physical parameters as such. Alternatively, the parameters derived by the parametrisation algorithm of the source finder can be used to derive reliability as a function of different source parameters, but it is important to note that for genuine sources those parameters can be affected by systematic errors and do not necessarily correspond to the original source parameters.
  
  The ideal source finder would produce a completeness and reliability of 100\%. In reality, however, we will have to find a compromise between good completeness and good reliability. In the case of \textsc{Du\-champ}, for example, decreasing the flux threshold for detections will lead to an increase in completeness, but at the cost of lower reliability.
  
  In our test of \textsc{Duchamp} on the set of 1024~point source models the software finds a total of 1103~sources of which 850 are genuine detections. The remaining 253~detections are false positives due to strong noise peaks in the data cube. These numbers translate into an overall completeness of 83.0\% and an overall reliability of 77.1\%.\footnote{Note that these numbers differ slightly from the ones quoted for run~5 in Figure~\ref{fig_tests} because a different realisation of the model was used in the initial tests. Reliability values will generally depend on the characteristics of the data cube under consideration (e.g.\ the size of the cube) and are therefore difficult to assess and compare.}
  
  \begin{figure*}[t]
    \centering
    \includegraphics[height=5.8cm]{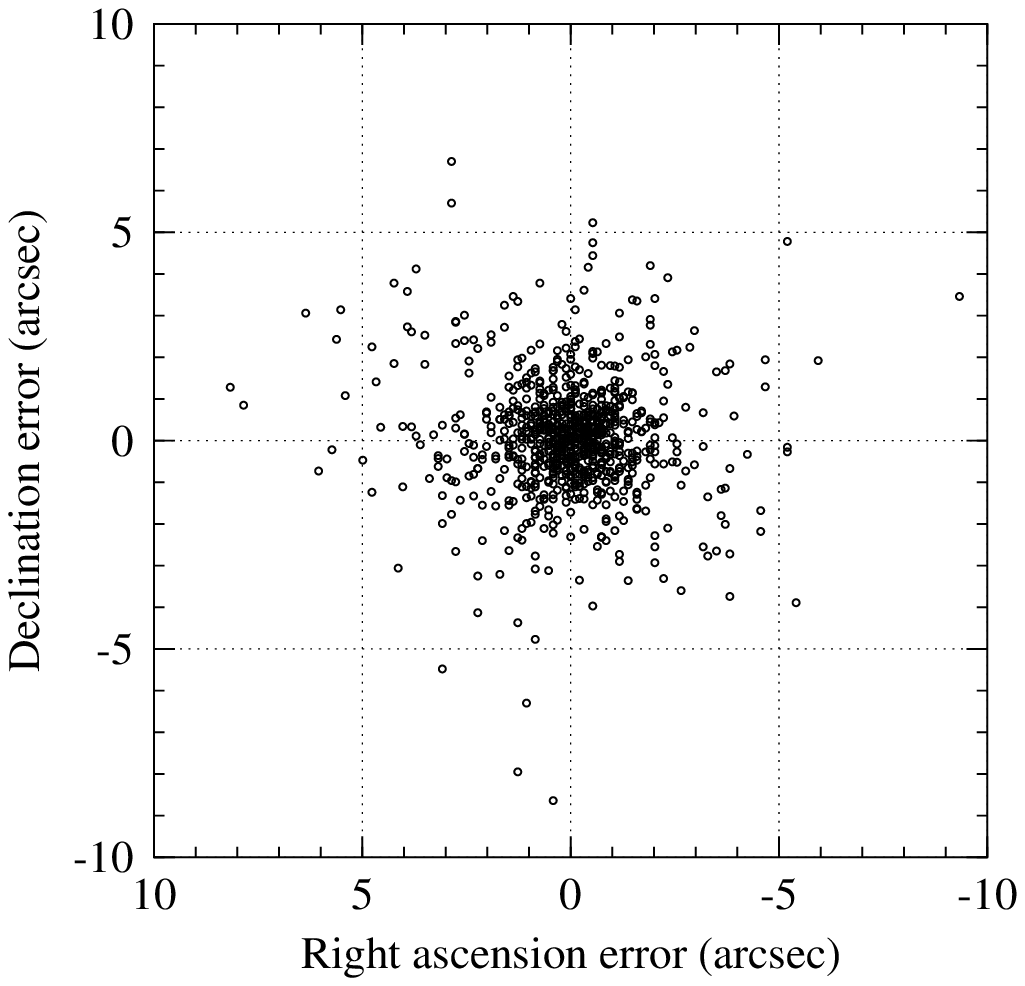}
    \hspace{0.5cm}
    \includegraphics[height=5.8cm]{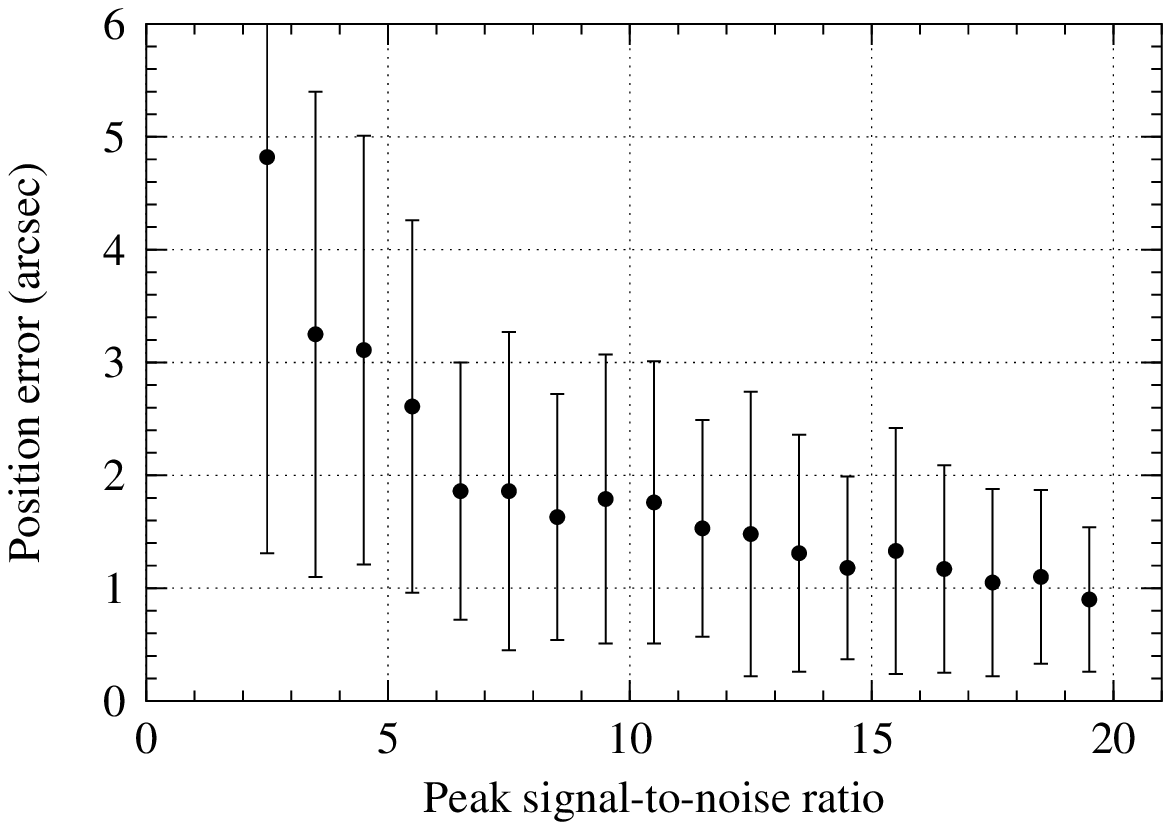}
    \caption{\textit{Left-hand panel:} Position error of the point source models in right ascension and declination. \textit{Right-hand panel:} Mean position error (black data points) and corresponding standard deviation (error bars) as a function of true peak signal-to-noise ratio in $1 \sigma$~bins.}
    \label{fig_position3d}
  \end{figure*}
  
  Completeness as a function of peak signal-to-noise ratio is plotted in the top panel of Figure~\ref{fig_completeness}. The detection list produced by \textsc{Duchamp} is complete down to a peak flux level of $F_{\rm peak} \approx 10 \sigma$, but below that level completeness decreases to below 50\% at about~$3 \sigma$. The completeness curve shows a much steeper rise when plotted against integrated flux instead of peak flux (middle panel of Figure~\ref{fig_completeness}). The 100\% completeness level is reached at $F_{\rm int} \approx 0.3~\mathrm{Jy \, km \, s}^{-1}$, corresponding to an \ion{H}{i}~mass of about $7 \times 10^{4}~\mathrm{M}_{\odot}$ at a distance of $1~\mathrm{Mpc}$, or $7 \times 10^{8}~\mathrm{M}_{\odot}$ at $100~\mathrm{Mpc}$, for the expected 8-hour integration per pointing of the WALLABY project on ASKAP. Below that flux level there is a sharp drop in completeness.
  
  \begin{figure*}[t]
    \centering
    \includegraphics[width=0.49\linewidth]{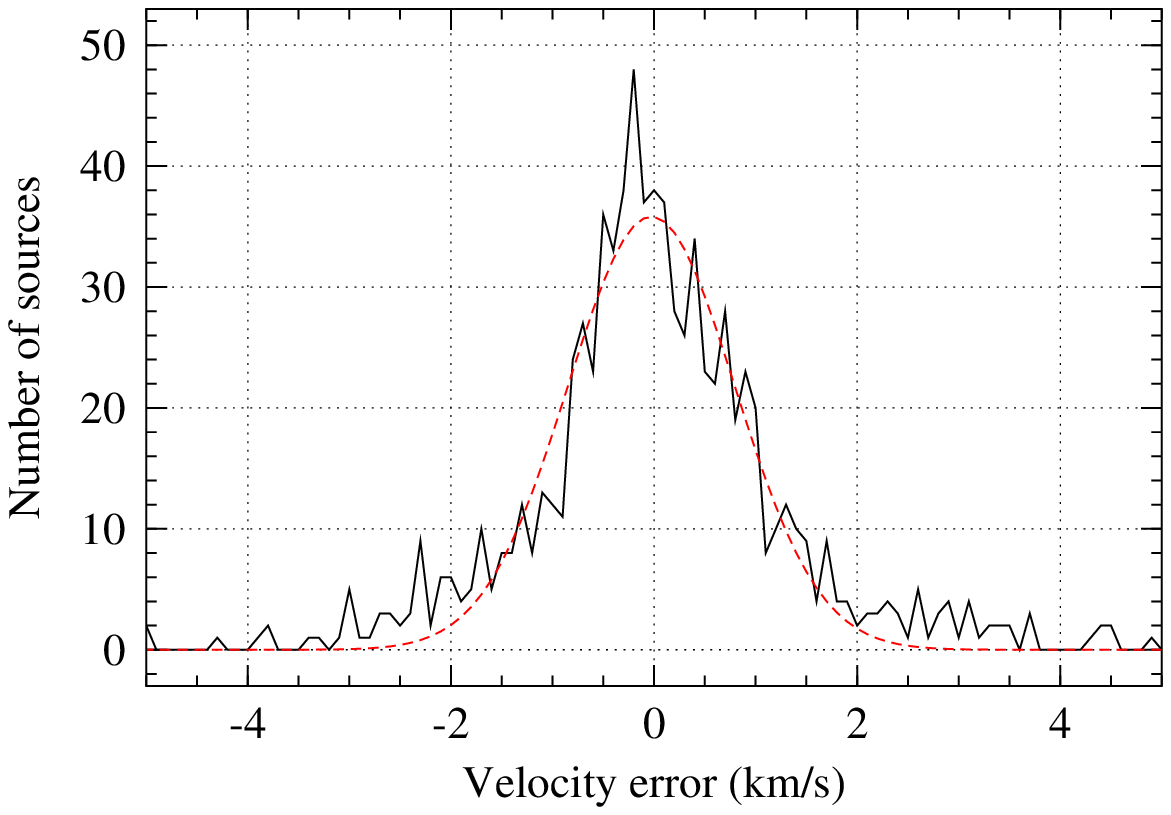}
    \hfill
    \includegraphics[width=0.49\linewidth]{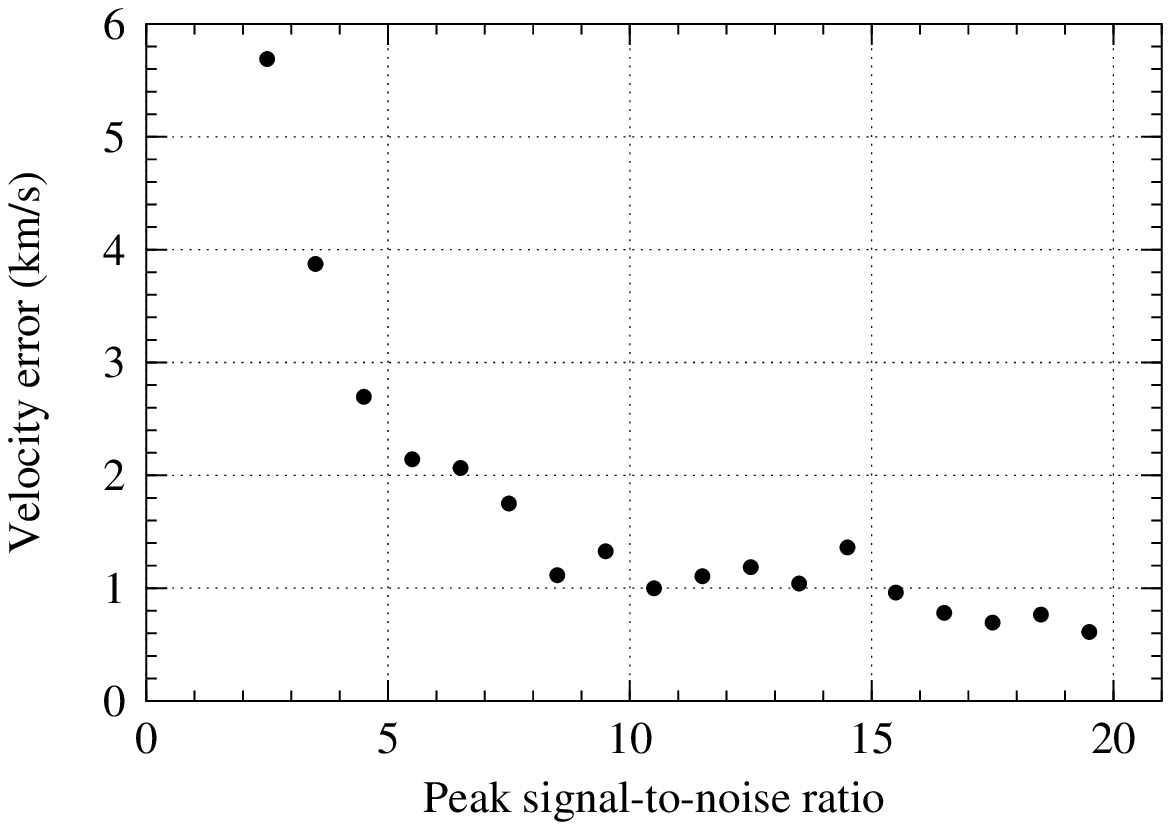}
    \caption{\textit{Left-hand panel:} Histogram of radial velocity errors (black curve) of the point source models in bins of $0.1~\mathrm{km \, s}^{-1}$. The red, dashed curve is the result of a Gaussian fit to the histogram. \textit{Right-hand panel:} Standard deviation of the velocity error of the sources as a function of true peak signal-to-noise ratio in bins of $1 \sigma$.}
    \label{fig_velocity}
  \end{figure*}
  
  The bottom panel of Figure~\ref{fig_completeness} shows completeness plotted as a function of true line width (FWHM), irrespective of the peak flux and integrated flux of a source. Over most of the covered line width range the completeness remains constant at approximately 90\%, but it gradually decreases to about 50\% below line widths of $10~\mathrm{km \, s}^{-1}$. This decrease is presumably the result of the `\`{a} trous' wavelet reconstruction of the data cube. By ignoring the smallest wavelet scales in the reconstruction we suppress the detection of noise peaks, but at the same time we are also less sensitive to genuine sources with narrow spectral lines.
  
  Reliability as a function of measured peak signal-to-noise ratio, measured integrated flux, and measured line width ($w_{50}$) is plotted in the top, middle, and bottom panels of Figure~\ref{fig_reliability}. \textsc{Duchamp} achieves 100\% reliability at a peak signal-to-noise ration of about~5 and an integrated flux level of approximately $0.1~\mathrm{Jy \, km \, s}^{-1}$. Reliabilities range between about 80\% to 100\% over most of the covered line width range, but drop significantly for sources with narrow lines of $w_{50} \lesssim 15~\mathrm{km \, s}^{-1}$ due to the increasing number of false detections associated with noise peaks. For line widths of less than about $5~\mathrm{km \, s}^{-1}$ the reliability increases sharply, because \textsc{Duchamp} effectively filters narrow signals caused by noise peaks through wavelet filtering and minimum channel requirements.
  
  However, as discussed earlier, reliability calculations are very difficult to assess and should be approached with great caution. First of all, reliability can only be specified as a function of \textit{measured} source parameters because false detections do not have genuine physical parameters. Any errors in a source finder's parametrisation will therefore affect the calculated reliability curves. Secondly, the actual reliability numbers are entirely meaningless in the case of model data as they depend on how the sources were distributed across the model data cube. Increasing the volume of the cube (without increasing the number of sources therein) will result in lower reliabilities as the fraction of false detections increases. Consequently, reliabilities can only be compared on a relative scale, e.g.~when testing different source finders on the same data set to determine which algorithm performs best.
  
  \subsubsection{Source Position}
  
  The resulting position errors are plotted in the left-hand panel of Figure~\ref{fig_position3d}. \textsc{Duchamp} does an excellent job in determining accurate source positions, with a mean position error of $0.0 \pm 1.6~\mathrm{arcsec}$ in right ascension and $0.1 \pm 1.5~\mathrm{arcsec}$ in declination.
  
  The mean position error (in terms of angular separation from the nominal source position) as a function of peak signal-to-noise ratio, in bins of $1 \sigma$, is shown in the right-hand panel of Figure~\ref{fig_position3d}. For bright sources of $F_{\rm peak} \approx 20 \sigma$ the mean position error is approximately $1~\mathrm{arcsec}$, increasing to about $5~\mathrm{arcsec}$ for $F_{\rm peak} \approx 3 \sigma$. These numbers correspond to only about 4\% and 19\%, respectively, of the FWHM of the synthesised beam.
  
  Two limitations should be noted at this point. First of all, in our models the source was always placed exactly on the central pixel of the data cube. We did not explicitly test placement of sources at positions in between the grid points of the cube, which---in the case of point sources---could result in reduced detection rates and less accurate source positions. Secondly, as with other source parameters, source positions will be inaccurate in cases where two or more sources are confused.

  \subsubsection{Radial Velocity}
  
  An overall histogram of radial velocity errors derived from the \textsc{Duchamp} run is shown in the left-hand panel of Figure~\ref{fig_velocity}. As expected, velocity errors have an approximately Gaussian distribution centred on zero. The mean velocity error of all sources is $0.0 \pm 1.7~\mathrm{km \, s}^{-1}$. The red, dashed curve in Figure~\ref{fig_velocity} shows the result of a Gaussian fit to the histogram. While the overall distribution of velocity errors follows the fitted Gaussian function, there are a few significant deviations, namely a somewhat higher and sharper peak in the centre (which is slightly shifted into the negative range) and conspicuous `wings' between $2$ and $3~\mathrm{km \, s}^{-1}$ (both positive and negative) where source counts are systematically too high with respect to the fit. The FWHM of the fitted Gaussian is $1.94 \pm 0.04~\mathrm{km \, s}^{-1}$, and the centroid is ${-0.026} \pm 0.017~\mathrm{km \, s}^{-1}$ which deviates from zero by about $1.5 \sigma$, reflecting the aforementioned negative offset of the peak of the histogram. These deviations from a pure Gaussian distribution are possibly caused by digitisation effects related to the segmentation of the frequency axis into discrete bins of $18.3~\mathrm{kHz}$ equivalent to $3.86~\mathrm{km \, s}^{-1}$.
  
  The standard deviation of the radial velocity error as a function of peak signal-to-noise ratio in $1 \sigma$ bins is shown in the right-hand panel of Figure~\ref{fig_velocity}. As expected, the standard deviation from the mean (which is essentially zero for all peak flux intervals) increases with decreasing peak flux. While for bright sources of $F_{\rm peak} \approx 20 \sigma$ the standard deviation is below $1~\mathrm{km \, s}^{-1}$, it increases to almost $6~\mathrm{km \, s}^{-1}$ for faint sources near the $3 \sigma$ level.

  \subsubsection{Line Width}
  
\begin{figure}[t]
    \centering
    \includegraphics[width=\linewidth]{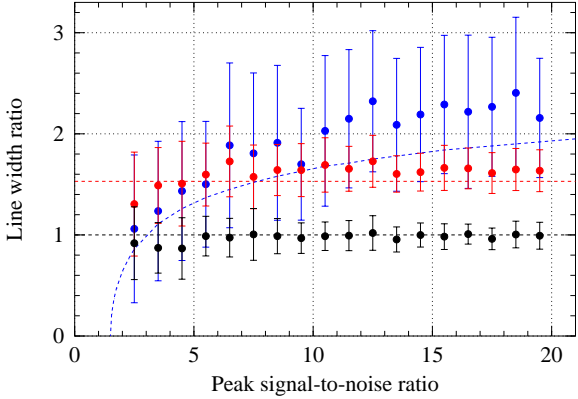}
    \caption{Ratio of measured versus true line width for the point source models as a function of true peak signal-to-noise ratio in bins of $1 \sigma$. The black data points show $w_{50}$, the red data points $w_{20}$, and the blue data points $w_{\rm vel}$ (for a $1.5 \sigma$ flux threshold), all of which have been divided by the original FWHM of the Gaussian line. The corresponding theoretical expectations for a Gaussian line profile are shown as the black, red, and blue dashed lines.}
    \label{fig_linewidth}
  \end{figure}

  Figure~\ref{fig_linewidth} shows the ratio of measured line width versus true line width (FWHM of the original Gaussian model) as a function of peak signal-to-noise ratio in bins of $1 \sigma$. \textsc{Duchamp} determines three different types of line width: $w_{50}$ is the full width at 50\% of the peak flux, $w_{20}$ is the full width at 20\% of the peak flux, and $w_{\rm vel}$ is the full detected line width of the source, i.e.~the width across all channels with detected flux. For a Gaussian line, $w_{50}$ is equivalent to the FWHM, and the ratio of $\mathrm{FWHM} / w_{50}$ should therefore be~$1$. The relation between $w_{20}$ and $w_{50}$ in the case of a Gaussian line is given by the constant factor of
  \begin{equation}
    \frac{w_{20}}{w_{50}} = 1.53 .
  \end{equation}
  Finally, the relation between $w_{\rm vel}$ and $w_{50}$, again assuming a Gaussian line profile, is defined via
%  \begin{equation}
%    \frac{w_{\rm vel}}{w_{50}} = \frac{\sqrt{{-8} \ln \left( \sigma / F_{\rm peak} \right) }}{2.35}
%  \end{equation}
  \begin{equation}
    \frac{w_{\rm vel}}{w_{50}} = \left[ \log_{\frac{1}{2}} \! \! \left( \! \frac{F_{\rm thr}}{F_{\rm peak}} \! \right) \right]^{\frac{1}{2}}
  \end{equation}
  where $F_{\rm thr} = n \times \sigma$ is the flux threshold used in the calculation of $w_{\rm vel}$. These theoretical relations are plotted in Figure~\ref{fig_linewidth} as the dashed lines for $w_{50}$ (black), $w_{20}$ (red), and $w_{\rm vel}$ (blue; for $F_{\rm thr} = 1.5 \sigma$).
  
  \textsc{Duchamp}'s measurement of $w_{50}$ (black data points) is in excellent agreement with the expectation (black, dashed line) over a wide range of peak signal-to-noise ratios. Only for faint sources of $F_{\rm peak} < 5 \sigma$ are the measured line widths on average slightly smaller than the true widths, but by no more than about 10 to 15\%.
  
  In contrast, \textsc{Duchamp}'s measurements of $w_{20}$ and $w_{\rm vel}$ (red and blue data points, respectively) are systematically too large over most of the covered range of signal-to-noise ratio as compared to the theoretical expectation (red and blue dashed lines, respectively). Only for faint sources of $F_{\rm peak} \lesssim 5 \sigma$ do the measured $w_{20}$ fall short of the theoretical ones. This result suggests that $w_{50}$ is the most accurate measurement of line width provided by \textsc{Duchamp} and should be used instead of $w_{20}$ and $w_{\rm vel}$ for the characterisation of astronomical sources. However, both $w_{50}$ and $w_{20}$ systematically fall short of the true line width for faint sources below $F_{\rm peak} \approx 5 \sigma$.
  
  \subsubsection{Peak Flux}
  
\begin{figure*}[ht]
    \centering
    \includegraphics[width=0.49\linewidth]{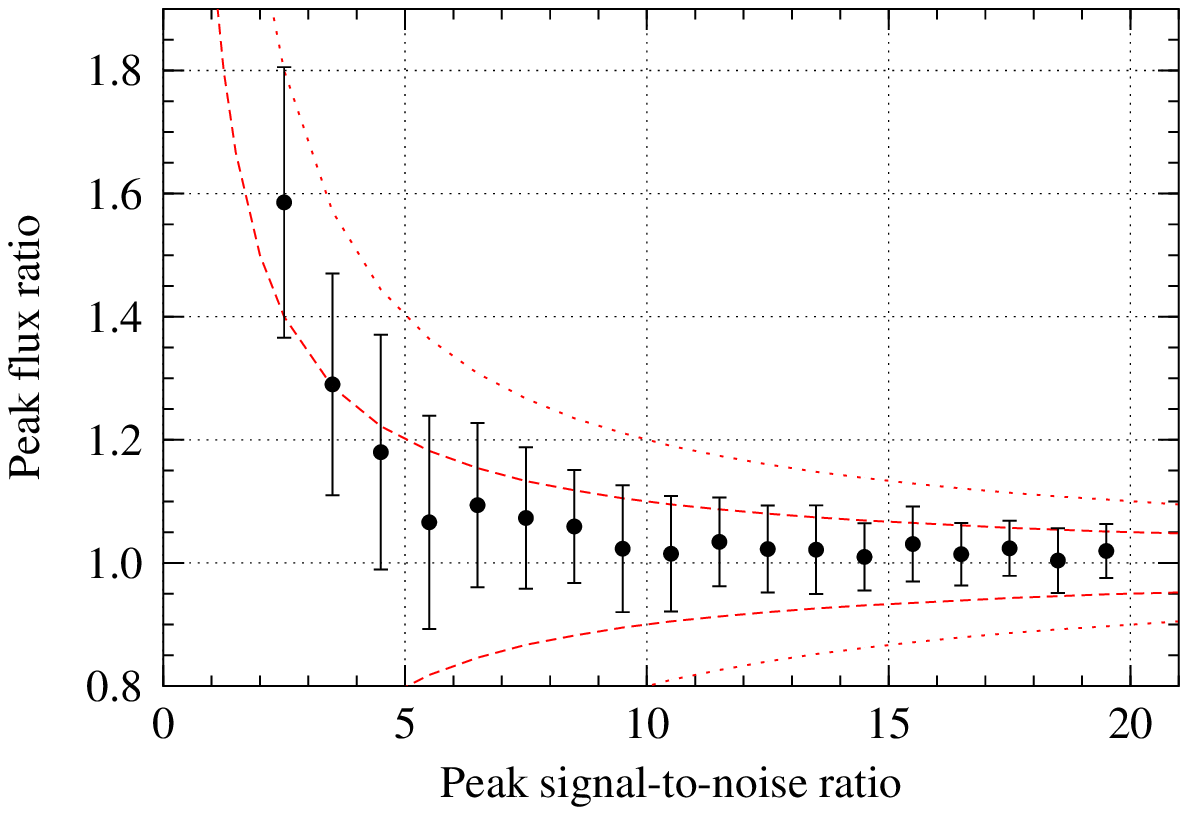}
    \hfill
    \includegraphics[width=0.49\linewidth]{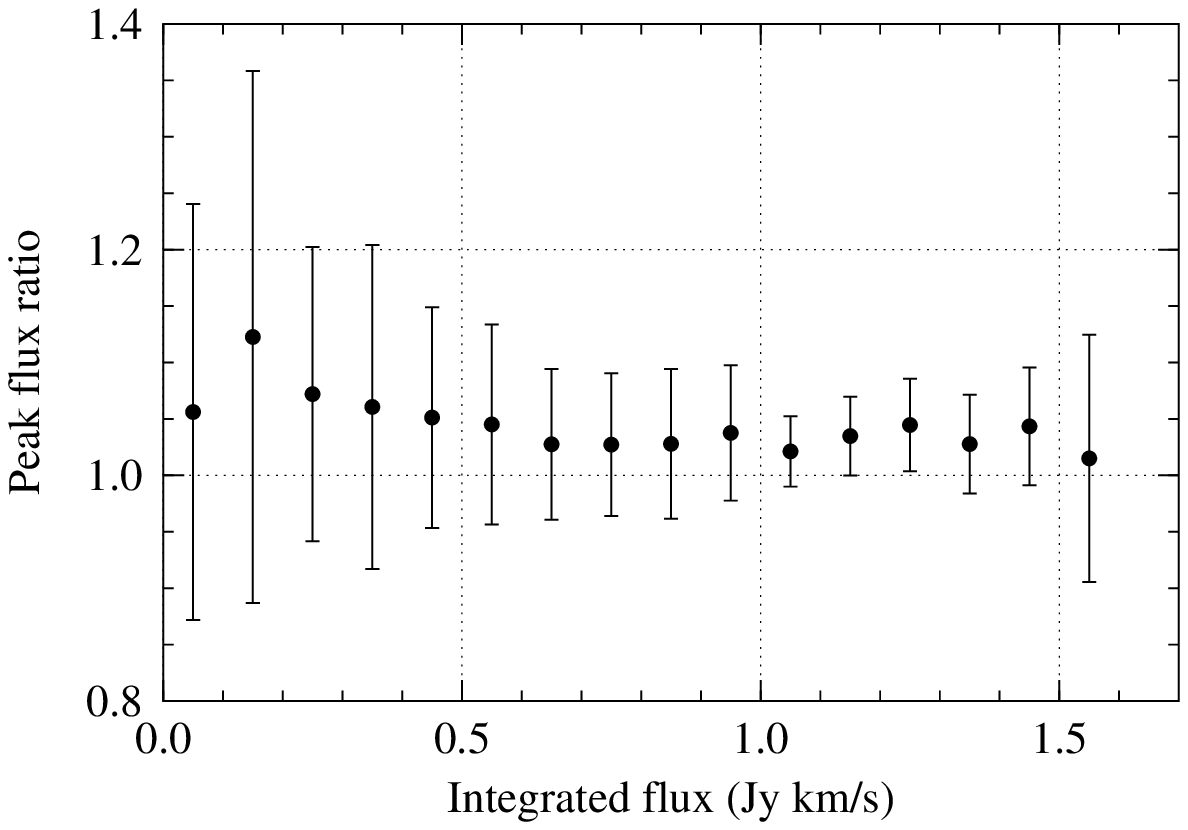}
    \caption{\textit{Left-hand panel:} Ratio of measured versus true peak flux (black data points) and corresponding standard deviation (error bars) of the model point sources as a function of true peak signal-to-noise ratio in $1 \sigma$ bins. The dashed and dotted red lines indicate the theoretical $\pm 1 \sigma$ and $\pm 2 \sigma$ envelopes, respectively. \textit{Right-hand panel:} Same, but as a function of true integrated flux in bins of $0.1~\mathrm{Jy \, km \, s}^{-1}$.}
    \label{fig_flux3d}
  \end{figure*}

  The ratio of recovered versus true peak flux of the model point sources is plotted in the left-hand panel of Figure~\ref{fig_flux3d} as a function of peak signal-to-noise ratio in $1 \sigma$ bins. The dashed and dotted red lines indicate the theoretical $\pm 1 \sigma$ and $\pm 2 \sigma$ envelopes, respectively. The right-hand panel shows the same figure, but as a function of integrated flux in bins of $0.1~\mathrm{Jy \, km \, s}^{-1}$. For bright sources of $F_{\rm peak} \gtrsim 10 \sigma$ \textsc{Duchamp} accurately recovers the peak flux of the sources, although there is the general tendency of measured peak fluxes being slightly too high on average. For fainter sources of $F_{\rm peak} \lesssim 5 \sigma$ there is a strong deviation, with measured fluxes being systematically too high by a significant factor. This is generally due to faint sources being more likely to be detected when their maximum coincides with a positive noise peak, whereas faint sources sitting on top of a negative noise peak will likely remain undetected, thereby creating a strong bias in the measurement of peak fluxes.
  
  Even for high peak signal-to-noise ratios the peak fluxes measured by \textsc{Duchamp} tend to be slightly too large. \textsc{Duchamp} determines the peak flux of a source by simply selecting the data element with the highest flux encountered. As mentioned before, this method is biased towards selecting data elements that have been affected by positive noise peaks. In sources with broad spectral signals there is a higher probability of finding a positive noise signal in one of the channels near the peak of the line that increases the signal beyond the actual line peak. This is a result of the source being well resolved in the spectral domain. Hence, peak fluxes measured by \textsc{Duchamp} will generally be too high irrespective of source brightness as long as the source is spectrally (or spatially) resolved. For very bright sources, however, the relative error will be negligible.
  
  \begin{figure*}[ht!]
    \centering
    \includegraphics[width=0.49\linewidth]{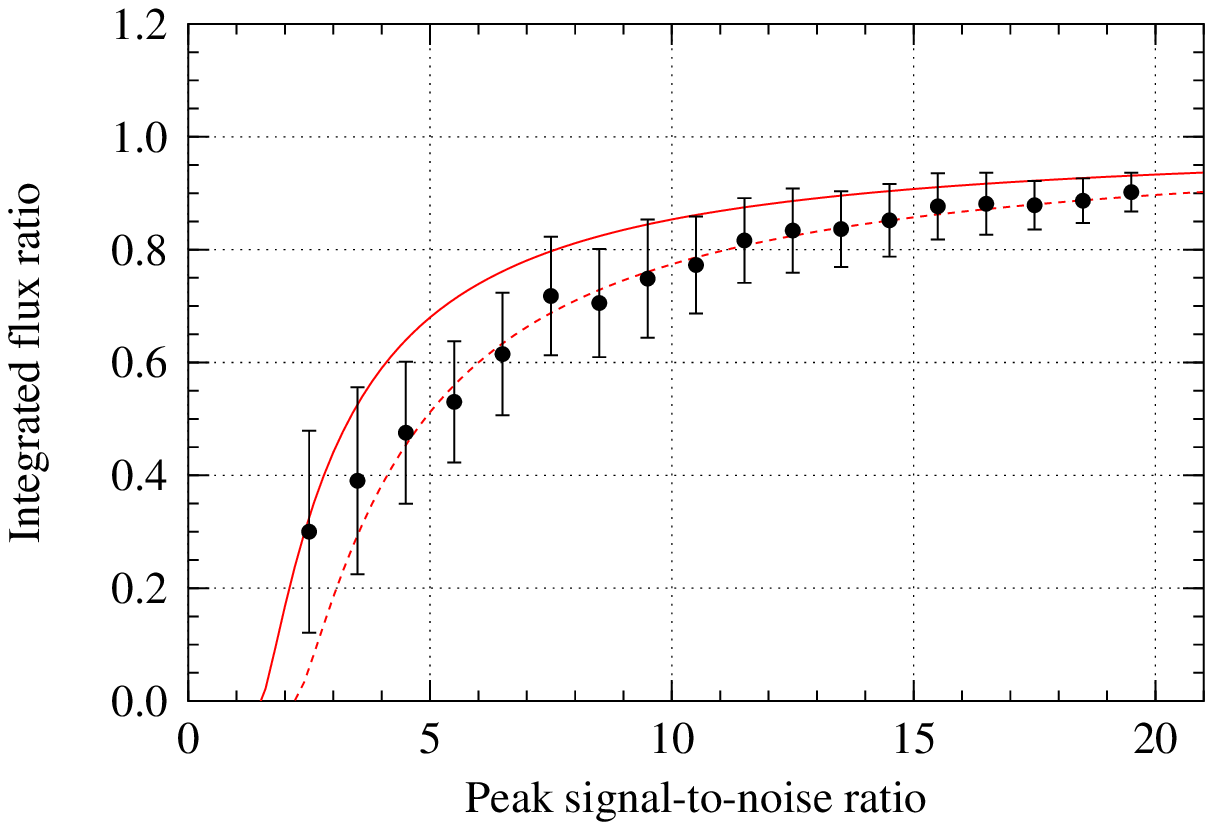}
    \hfill
    \includegraphics[width=0.49\linewidth]{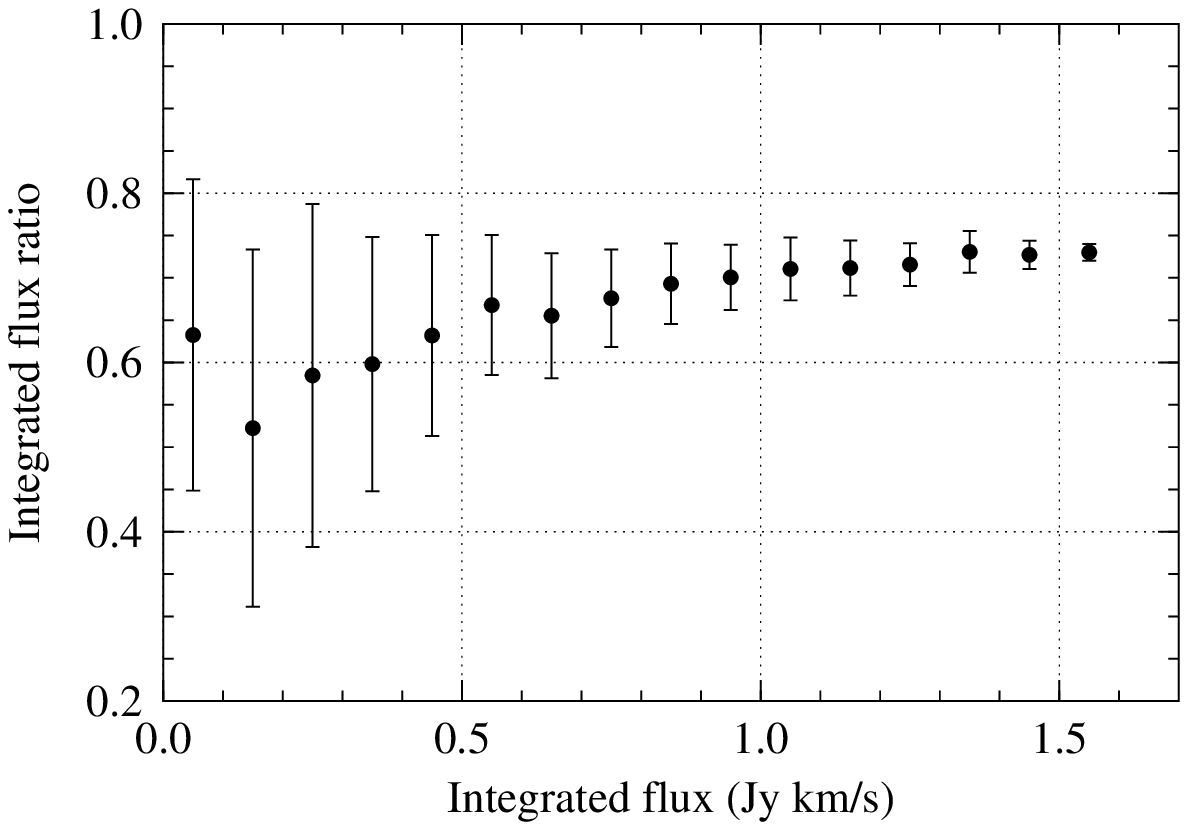}
    \caption{\textit{Left-hand panel:} Ratio of measured versus true integrated flux (black data points) and corresponding standard deviation (error bars) for the model point sources as a function of true peak signal-to-noise ratio in bins of $1 \sigma$. The solid red curve shows the theoretical expectation for the $1.5 \sigma$ flux threshold used in our test. The dotted red curve shows the best fit to the data points, corresponding to an effective flux threshold of $2.2 \sigma$. \textit{Right-hand panel:} Same, but as a function of true integrated flux in bins of $0.1~\mathrm{Jy \, km \, s}^{-1}$.}
    \label{fig_intflux}
  \end{figure*}

  \subsubsection{Integrated Flux}
  
  The ratio of measured versus true integrated flux of the model point sources as a function of peak signal-to-noise ratio in bins of $1 \sigma$ is presented in the left-hand panel of Figure~\ref{fig_intflux}. The right-hand panel shows the same figure, but as a function of integrated flux in bins of $0.1~\mathrm{Jy \, km \, s}^{-1}$. Apparently, \textsc{Duchamp}'s measurement of the integrated flux of a source is systematically too low by a significant factor. Even for the brightest sources of $F_{\rm peak} \approx 20 \sigma$ only about 90\% of the true flux is recovered by \textsc{Duchamp}, and that figure drops to well below 50\% for faint sources of $F_{\rm peak} < 5 \sigma$.
  
  This issue is likely caused by the fact that \textsc{Duchamp} only considers pixels above the detection threshold when calculating the integrated flux. Pixels below the threshold, while potentially contributing significantly to the overall flux of a source, are not included in the summation carried out by \textsc{Duchamp}, resulting in integrated fluxes being systematically too small.

  In order to study the expected decrease in the integrated flux measurement, let us assume a point source with Gaussian line profile being observed with a telescope with radially symmetric Gaussian point spread function (PSF),
  \begin{equation}
    F(x,y,v) = F_{\rm peak} \exp \! \left( {-\frac{x^{2} + y^{2}}{2 \sigma_{\rm PSF}^{2}}} - \frac{v^{2}}{2 \sigma_{v}^{2}} \right) ,
  \end{equation}
  with amplitude, $F_{\rm peak}$, velocity dispersion, $\sigma_{v}$, and PSF size, $\sigma_{\rm PSF}$. The integrated flux measurement can then be considered as the integral under the three-dimensional Gaussian brightness profile across the frequency/velocity range, $\pm v_{0}$, and the spatial range, $\pm x_{0}$ and $\pm y_{0}$, over which the flux of the line is above the detection threshold, thus
  \begin{eqnarray}
    F_{\rm int} & = &  \int \limits_{-x_{0}}^{~x_{0}} \int \limits_{-y_{0}}^{~y_{0}} \int \limits_{-v_{0}}^{~v_{0}} \! \! F(x,y,v) \, \mathrm{d}x \, \mathrm{d}y \, \mathrm{d}v \\
                & = & F_{\rm peak} (2 \uppi)^{3/2} \sigma_{\rm PSF}^{2} \sigma_{v} \, \mathrm{erf} \! \left( \! \frac{x_{0}}{\sqrt{2} \sigma_{\rm PSF}} \! \right) \nonumber \\
                &   & \times \: \mathrm{erf} \! \left( \! \frac{y_{0}}{\sqrt{2} \sigma_{\rm PSF}} \! \right) \mathrm{erf} \! \left( \! \frac{v_{0}}{\sqrt{2} \sigma_{v}} \! \right) , \label{eqn_flux}
  \end{eqnarray}
  where $\mathrm{erf}(x)$ is the error function. Inserting the appropriate integration limits and then dividing Equation~\ref{eqn_flux} by the total flux (i.e.~integrated over $\pm \infty$) leads to a theoretical integrated flux ratio of
  \begin{equation}
    \frac{F_{\rm int}}{F_{\rm tot}} = \left[ \mathrm{erf} \left( \sqrt{{-\ln} \left( F_{\rm thr} / F_{\rm peak} \right) } \right) \right]^{3} \label{eqn_flux2}
  \end{equation}
  with a flux threshold of $F_{\rm thr} = n \times \sigma$.
  
  The resulting theoretical integrated flux according to Equation~\ref{eqn_flux2}, assuming a $1.5 \sigma$~threshold, is shown as the solid red curve in Figure~\ref{fig_intflux}. The integrated fluxes measured by \textsc{Duchamp} are only slightly below what one would expect from a simple integration over a three-dimensional Gaussian. A fit to the data points instead yields an effective flux threshold of $2.2 \sigma$, shown as the dotted red curve in Figure~\ref{fig_intflux}, which is slightly larger than the $1.5 \sigma$ used when running \textsc{Duchamp}. It is not quite clear why \textsc{Duchamp} performs worse than expected. The discrepancy could be due to the fact that the software sums over discrete pixels whereas we assumed continuous integration in our mathematical model. This will likely result in small differences, particularly in those cases where the number of elements across the Gaussian profile is small. In our case, as we are dealing with point sources, this is certainly true for the spatial dimension.
  
  In summary, integrated flux measurements provided by \textsc{Duchamp} are systematically too small and will need to be corrected substantially to compensate for the systematic offset.

  \begin{figure*}[t]
    \begin{center}
      \includegraphics[width=\linewidth]{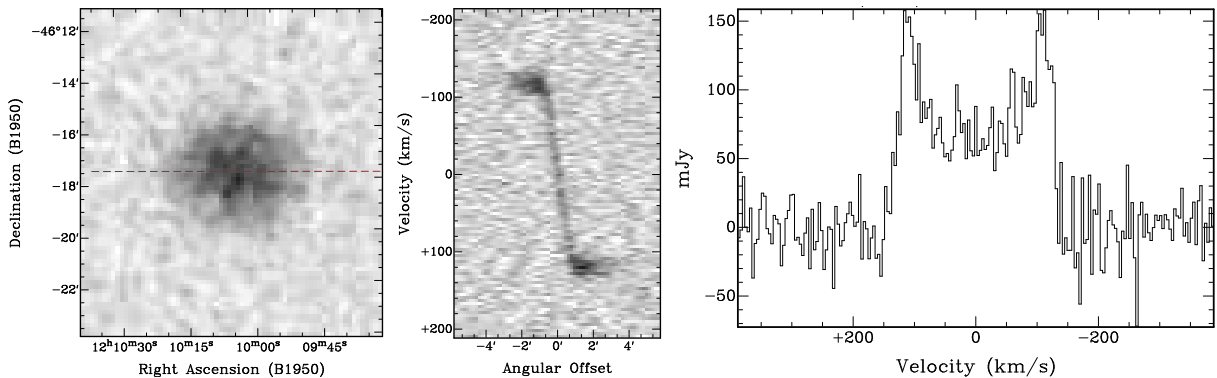}
      \caption{Example of a model galaxy generated for testing \textsc{Duchamp}. The left-hand panel shows the zeroth moment of the model, the middle panel shows the position-velocity diagram along the dashed, red line, and the right-hand panel depicts the integrated spectrum of the model galaxy.}
      \label{fig_maps_2}
    \end{center}
  \end{figure*}

  \section{Models of Disc Galaxies}
  \label{sect_galaxies}
  
  In order to test the performance of \textsc{Duchamp} on more realistic, extended sources, we generated 1024~artificial \ion{H}{i} models of galaxies with a wide range of parameters, using a programme written in~C for direct manipulation of FITS data cubes. All galaxies were modelled as infinitely thin discs with varying inclination ($0^{\circ}$ to $89^{\circ}$), position angle ($0^{\circ}$ to $180^{\circ}$), and rotation velocity ($20$ to $300~\mathrm{km \, s}^{-1}$). For any one galaxy, inclination and position angle were considered to be constant over the radial extent of the disc, while the rotation velocity increases linearly from $0$ to $v_{\rm rot}$ between the centre and $0.5$~times the semi-major axis of the disc and remains constant beyond that radius. Individual spectral profiles across the disc were assumed to be Gaussian with a dispersion of $9.65~\mathrm{km \, s}^{-1}$ (equivalent to $2.5$~times the width of a spectral channel). The radial surface brightness profile was assumed to be Gaussian, too, resulting in an elliptical Gaussian brightness distribution on the sky.
  
  Next, we again generated an artificial ASKAP visibility data set of pure Gaussian noise at a frequency of $1.4~\mathrm{GHz}$ with characteristics similar to the WALLABY survey, using the \textsc{Miriad} task \textsc{uvgen} with parameters as listed in Table~\ref{tab_model2}. The visibility data were Fourier-transformed to create an image of the point spread function and a noise data cube with $1601 \times 1601$~spatial pixels of $10~\mathrm{arcsec}$ size and $201$~spectral channels. We then convolved the model galaxies with a clean beam derived from fitting a Gaussian to the central peak of the point spread function. Finally, the convolved galaxy models were placed on a regular grid of $32 \times 32$~galaxies and added to the noise cube to create the final data cube of model galaxies for the testing of \textsc{Duchamp}.
  
  The moment-zero map, position-velocity map, and integrated spectrum of one of the model galaxies is shown in Figure~\ref{fig_maps_2} for illustration. As with the point sources, all galaxies were centred on a pixel, although for extended sources we do not expect any significant effect from shifting the source centre with respect to the pixel centre. Again, all sources are isolated, and we did not attempt to test \textsc{Duchamp} in a situation where source crowding occurs.
  
  \begin{table}[hb!]
    \begin{center}
      \caption{Summary of the parameters used to generate the visibility data set and noise image for the galaxy models.}
      \label{tab_model2}
      \begin{tabular}{lrl}
        \hline
        Parameter (visibility) &         Value & Unit           \\
        \hline
        Number of antennas     &          $36$ &                \\
        System temperature     &          $50$ & $\mathrm{K}$   \\
        Declination            & $-45^{\circ}$ &                \\
        Total integration time &           $8$ & $\mathrm{h}$   \\
        Hour angle range       &       $\pm 4$ & $\mathrm{h}$   \\
        Cycle time             &          $36$ & $\mathrm{s}$   \\
        Stokes parameters      &             I &                \\
        Number of channels     &         $201$ &                \\
        Frequency              &        $1.42$ & $\mathrm{GHz}$ \\
        Channel width          &       $18.31$ & $\mathrm{kHz}$ \\
                               &        $3.86$ & $\mathrm{km \, s}^{-1}$ \\
        \hline
        Parameter (image)      &            Value & Unit                 \\
        \hline
        Final image size       & $1601 \times 1601$ & $\mathrm{px}$        \\
        Field diameter         &     $4\fdeg{}45$ &                      \\
        Pixel size             &             $10$ & $\mathrm{arcsec}$    \\
        Robustness             &              $0$ &                      \\
        Gaussian $uv$ taper    &           $7.28$ & $\mathrm{k} \lambda$ \\
                               &           $1.54$ & $\mathrm{km}$        \\
        \textsc{rms} noise     &           $1.86$ & $\mathrm{mJy}$       \\
        Synthesised beam       &                  &                      \\
        \quad major axis       &           $30.9$ & $\mathrm{arcsec}$    \\
        \quad minor axis       &           $30.5$ & $\mathrm{arcsec}$    \\
        \quad position angle   &     $50\fdeg{}8$ &                      \\
        \hline
      \end{tabular}
    \end{center}
  \end{table}
  
  It is important to note at this point that the resulting model galaxies, while exhibiting some of the spatial and spectral characteristics of real spiral galaxies, have been simplified to a great extent, resulting in limitations that need to be kept in mind when interpreting the results presented in this section. Firstly, the assumption of an infinitely thin disc will result in unrealistic edge-on galaxies, with integrated fluxes as well as individual spectral line widths across the disc being too small. Secondly, parameters such as peak flux, angular size, or rotation velocity were all varied independently of each other, resulting in unrealistic combinations of galaxy parameters in some cases. The purpose of the models is to cover a vast parameter range of extended sources irrespective of whether that entire range is populated by real galaxies. Even if disc galaxies with a certain combination of parameters do not exist, other objects, such as irregular galaxies or high-velocity clouds, could still cover those regions of parameter space, and their exploration will therefore be meaningful.
  
  \begin{table*}[t]
    \begin{center}
      \caption{\textsc{Duchamp} input parameters explicitly set in the input parameter file for the galaxy models. The default values of \textsc{Duchamp} were used for all other parameters.}
      \label{tab_input2}
      \begin{tabular}{lrrrl}
        \hline
        Parameter             &   Run~1 &   Run~2 &   Run~3 & Comment \\
        \hline
        threshold             & 0.00186 & 0.00186 & 0.00186 & 1.0 $\times$ \textsc{rms} \\
        minPix                &      10 &      10 &      10 & \\
        minChannels           &       5 &       5 &       3 & \\
        flagAdjacent          &    true &    true &    true & \\
        flagGrowth            &   false &    true &    true & \\
        growthThreshold       &      -- & 0.00093 & 0.00093 & 0.5 $\times$ \textsc{rms} \\
        flagRejectBeforeMerge &   false &    true &    true & \\
        flagATrous            &    true &    true &    true & Wavelet reconstruction \\
        reconDim              &       3 &       3 &       3 & in 3 dimensions \\
        snrRecon              &       2 &       2 &       2 & \\
        scaleMin              &       3 &       3 &       3 & \\
        \hline
      \end{tabular}
    \end{center}
  \end{table*}
  
  \subsection{Running {\sc Duchamp}}
  \label{sect_galaxies_duchamp}
  
  We ran \textsc{Duchamp} (version~1.1.12) on the model galaxy cube several times with slightly different input parameters to compare the performance. The different input parameters explicitly set in the parameter file are listed and compared in Table~\ref{tab_input2}. In all cases we employed a $1 \sigma$ flux threshold, equivalent to about $1.9~\mathrm{mJy}$, and performed a three-dimensional `\`a{}~trous' wavelet reconstruction with a minimum scale of~3 and a flux threshold of $2 \sigma$ for wavelet components to be included in the reconstructed cube. The slightly larger minimum scale as compared to the point source models is motivated by the fact that we are now dealing with spatially and spectrally much more extended sources. In addition, we varied the number of contiguous spectral channels required for detections and used \textsc{Duchamp}'s growth criterion in a few runs with a growth flux threshold of $0.5 \sigma$. The latter method will grow detections to flux levels below the original detection threshold, resulting in more accurate source parametrisation. As it turned out, the change from 5 to 3~consecutive spectral channels for detections (run~2 versus~3) did not have any major impact on the results. Hence, only the results of runs~1 and~3 will be presented and discussed here.
  
  \begin{figure}[t!]
    \begin{center}
      \includegraphics[width=\linewidth]{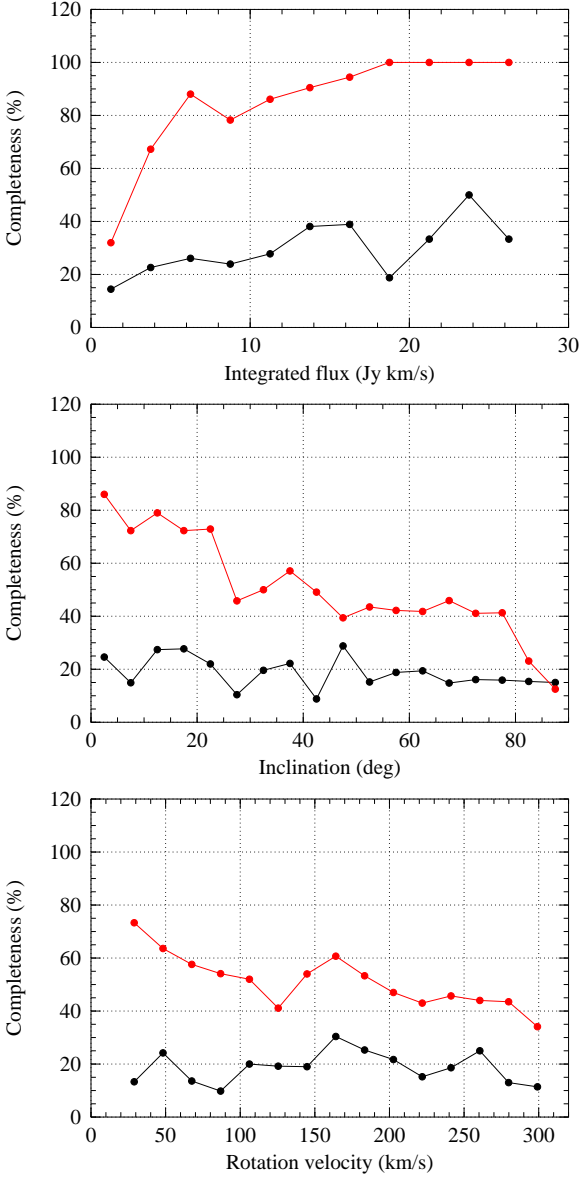}
      \caption{Completeness for the galaxy models as a function of true integrated flux in bins of $2.5~\mathrm{Jy \, km \, s}^{-1}$ (top panel), galaxy inclination in bins of $5^{\circ}$ (middle panel), and rotation velocity in bins of $19.3~\mathrm{km \, s}^{-1}$ (bottom panel) for runs~1 (black) and~3 (red).}
      \label{fig_completeness2}
    \end{center}
  \end{figure}
  
  \begin{figure}[t!]
    \begin{center}
      \includegraphics[width=\linewidth]{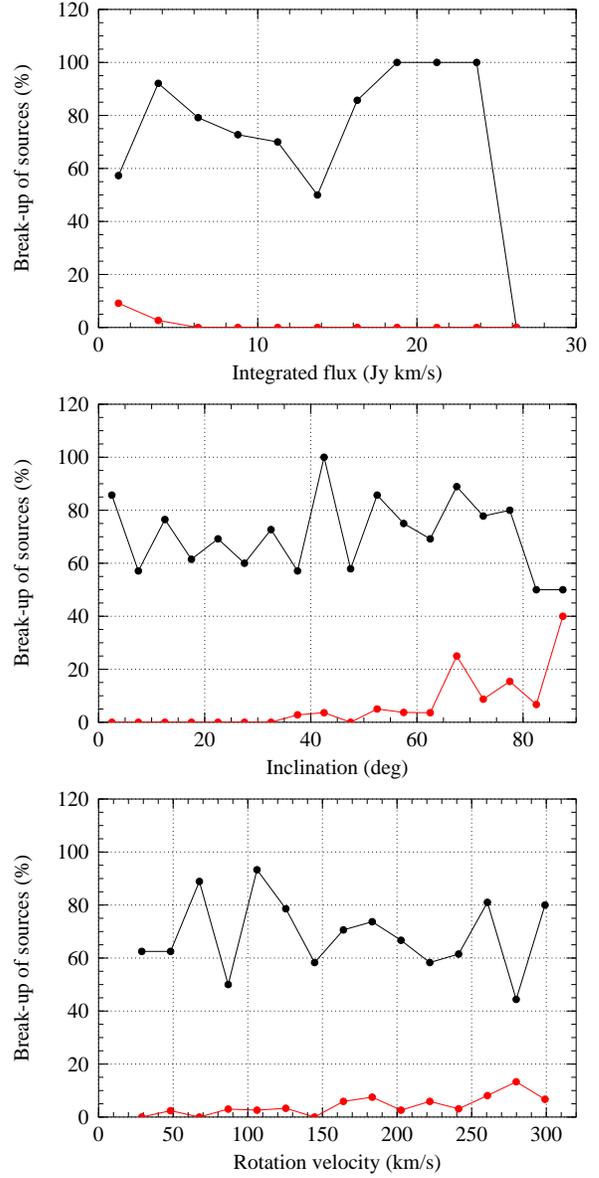}
      \caption{Fraction of model galaxies being broken up into two or more separate detections by \textsc{Duchamp} as a function of true integrated flux in bins of $2.5~\mathrm{Jy \, km \, s}^{-1}$ (top panel), galaxy inclination in bins of $5^{\circ}$ (middle panel), and rotation velocity in bins of $19.3~\mathrm{km \, s}^{-1}$ (bottom panel) for runs~1 (black) and~3 (red).}
      \label{fig_breakup}
    \end{center}
  \end{figure}
  
  In order to compare the outcome of \textsc{Duchamp} with the original input catalogue, we wrote a short Python script that reads in and processes the different catalogues. The script first reads in the \textsc{Duchamp} output catalogue, the original model catalogue, and a special mask data cube marking pixels with emission in the original model by assigning them a unique number characteristic to each input source. The script then cycles through all the detections made by \textsc{Duchamp} and decides for each detection whether it is genuine or not by checking the value of the mask data cube at the same position. If the detection is found to be genuine, the script will cycle through the original model catalogue to extract the actual input parameters of the respective source for comparison with the parametrisation results of \textsc{Duchamp}.
  
  At the end of this process we get a match of detected sources with original input sources, allowing us to calculate parameters such as completeness, reliability, and the fraction of sources being broken up into multiple detections by \textsc{Duchamp}. In addition, we are able to compare the original parameters of each source with those determined by \textsc{Duchamp} to test the performance of \textsc{Duchamp}'s parametrisation algorithms.
  
  \subsection{Results}
  
  \subsubsection{Completeness and Reliability}
  
  For run~1 (without growing of detections to flux levels below the threshold), $436$ out of $1063$~detected sources are genuine, resulting in an overall reliability of $41\%$. As many original sources got broken up into multiple detections, only $194$ of the $1024$~input galaxies were detected, yielding an overall completeness of only~$19\%$. There is a significant improvement for run~3 (with growing of detections to a flux level of $0.5 \sigma$), where $542$ out of $1051$~detected sources are genuine (reliability of $52\%$), but this time $521$ of the $1024$~input galaxies were detected, resulting in a much improved overall completeness of~$51\%$.
  
  Completeness as a function of different galaxy parameters is shown in Figure~\ref{fig_completeness2} for runs~1 and~3 (black and red data points, respectively). As mentioned before, run~1 resulted in very low completeness values of typically only about 20\% and no strong variation with either the integrated flux of a source or its inclination and rotation velocity. By growing detections to a flux level of $0.5 \sigma$ (run~3) we achieved much higher completeness levels over a large parameter range. 100\% completeness is achieved for sources of $F_{\rm int} \gtrsim 20~\mathrm{Jy \, km \, s}^{-1}$, and completeness levels reach 50\% at $F_{\rm int} \approx 2.5~\mathrm{Jy \, km \, s}^{-1}$. The latter corresponds to an \ion{H}{i}~mass sensitivity of $6 \times 10^{5}~\mathrm{M}_{\odot}$ at a distance of $1~\mathrm{Mpc}$, or $6 \times 10^{9}~\mathrm{M}_{\odot}$ at $100~\mathrm{Mpc}$, for the expected 8-hour integration per pointing of the WALLABY project on ASKAP.
  
  As shown in the middle and bottom panels of Figure~\ref{fig_completeness2}, there is a strong variation of completeness with both inclination and rotation velocity of the galaxies. While face-on galaxies are on average detected at completeness levels near 80\%, \textsc{Duchamp} struggles to find edge-on galaxies, yielding average completeness levels of only about 20\% for galaxies with inclination angles greater than $80^{\circ}$. This effect is caused by the combination of two separate effects. Firstly, as a result of the limitations from our assumption of an infinitely thin disc, edge-on galaxies have typically lower integrated fluxes than face-on galaxies. Secondly, edge-on galaxies typically have a broader spectral signature as a result of their higher projected rotation velocity, making it more difficult for \textsc{Duchamp} to pick up their extended signal.
  
  The latter effect can also be seen in the bottom panel of Figure~\ref{fig_completeness2}, where completeness levels systematically decrease as a function of increasing rotation velocity of a galaxy, irrespective of its inclination or integrated flux, confirming that on average \textsc{Duchamp} is more likely to pick up face-on galaxies with narrow spectral lines. It is important to note, however, that at a given distance galaxies with higher rotation velocity will typically have a larger \ion{H}{i}~mass and are therefore more likely to be detected than galaxies with lower rotation velocity at the same distance.
  
  \begin{figure*}[t]
    \begin{center}
      \includegraphics[width=0.9\linewidth]{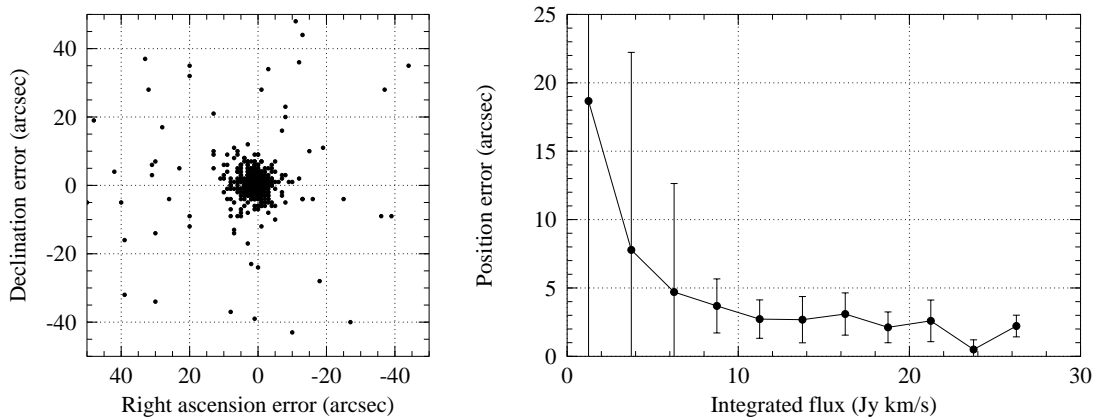}
      \caption{\textit{Left-hand panel:} Position errors (from run~3) for the model galaxies in right ascension and declination. \textit{Right-hand panel:} Mean absolute position error (data points) and standard deviation (error bars) as a function of true integrated flux in bins of $2.5~\mathrm{Jy \, km \, s}^{-1}$.}
      \label{fig_position2}
    \end{center}
  \end{figure*}
  
  \subsubsection{Break-up of Sources into Multiple Components}
  
  Due to their rotation velocity, spiral galaxies often exhibit a large radial velocity gradient across their projected disc on the sky, resulting in the characteristic double-horn profile of their integrated spectrum. This, however, can result in the two halves of a galaxy being detected as two separate sources by \textsc{Duchamp}, in particular in the case of faint, edge-on galaxies with large rotation velocities.
  
  In Figure~\ref{fig_breakup} we have plotted the fraction of detected model galaxies that were broken up into two or more separate detections by \textsc{Duchamp} as a function of true integrated flux (top panel), inclination (middle panel), and rotation velocity (bottom panel). For run~1 (black data points) there is a very high fraction of multiple detections, typically about 60--80\%, with no strong variation with either integrated flux of the galaxy or inclination and rotation velocity of the disc. In total, $136$ out of the $194$~detected galaxies, or $70.1\%$, were broken up into multiple components by \textsc{Duchamp}.
  
  Growing detections to the $0.5 \sigma$ level (run~3, red data points) results in a major improvement, with the number of multiple detections (in total $21$ out of $521$~detected galaxies, or $4.0\%$) dropping to zero over most of the covered parameter range. Only for faint sources of $F_{\rm int} \lesssim 5~\mathrm{Jy \, km \, s}^{-1}$ does the fraction of multiple detections gradually increase up to about 10\% at the low end of the flux spectrum. Figure~\ref{fig_breakup} also clearly shows the expected increase in multiple detections for galaxies of higher inclination ($i \gtrsim 40^{\circ}$) and rotation velocity ($v_{\rm rot} \gtrsim 150~\mathrm{km \, s}^{-1}$), which is the result of the double-horn profile becoming wider and more pronounced as the radial velocity gradient in the plane of the sky increases.
  
  A similar case, although more difficult to assess, is the detection of only one half of a galaxy (one horn of the double-horn profile), whereas the other half remains undetected. As there is only a single detection of each affected galaxy, such partial detections are much more difficult to identify. They should, however, result in a significant offset of both the measured position and radial velocity of the detected source with respect to the location of the originating galaxy.
  
  In the case of run~3, $62$~out of $500$~single detections show velocity errors of more than $20~\mathrm{km \, s}^{-1}$, with~$28$ even exceeding $150~\mathrm{km \, s}^{-1}$. The former corresponds to a fraction of $12.4\%$ of all single detections. Similarly, $62$~out of $500$~singly detected sources have a position error of more than $20~\mathrm{arcsec}$, which again corresponds to a fraction of $12.4\%$.\footnote{There is no exact match between the $62$~sources with large position error and the $62$~sources with large velocity error. A total of $66$~sources fulfil either of the two criteria.}
  
  These results suggest that, even when growing detections down to the $0.5 \sigma$ level, there is a significant number of partial (approximately $66$~sources) or multiple ($21$~sources) detections, corresponding to an overall fraction of about $16.7\%$ of all genuine detections. Such cases need to be identified in the output catalogue produced by \textsc{Duchamp}, as otherwise they will introduce a significant bias in the measurement of source parameters such as line width and \ion{H}{i}~mass. Identification of broken-up sources will be a very difficult task in practice, as it may be impossible to decide whether two detections are part of the same source or two separate sources in close proximity. While the growing of detections to lower flux levels can in principle reduce the fraction of sources being broken up, an undesirable side effect will be the potential merging of neighbouring sources, e.g.~close galaxy pairs in group or cluster environments.
  
  \subsubsection{Source Position}
  
  The left-hand panel of Figure~\ref{fig_position2} shows a scatter plot of position errors for the model galaxies (based on run~3) in right ascension and declination. The mean position errors in right ascension and declination are $1.7 \pm 14.7~\mathrm{arcsec}$ and $0.6 \pm 12.7~\mathrm{arcsec}$, respectively. The standard deviation is fairly large because there are several sources with position errors of tens of arcsec, well beyond the central concentration in the plot. These are cases in which only one half of a galaxy was detected as a source, whereas the other half remained undetected, resulting in systematic offsets in position as well as velocity with respect to the original model.
  
  When excluding such cases of partial detections by only considering detections with position errors of less than $15~\mathrm{arcsec}$ in both right ascension and declination, we obtain corrected errors of $0.9 \pm 3.6~\mathrm{arcsec}$ in right ascension and $0.5 \pm 3.6~\mathrm{arcsec}$ in declination.
  
  The combined, absolute position error as a function of true integrated flux is shown in the right-hand panel of Figure~\ref{fig_position2}. For bright sources of $F_{\rm int} \gtrsim 10~\mathrm{Jy \, km \, s}^{-1}$ source positions are very accurate with typical errors of about $2.5~\mathrm{arcsec}$. Towards the faint end of the diagram both mean error and standard deviation increase substantially, partly as a result of increasing statistical uncertainties, but also due to an increasing fraction of galaxies that are only partially detected.
  
  \begin{figure}[t]
    \begin{center}
      \includegraphics[width=\linewidth]{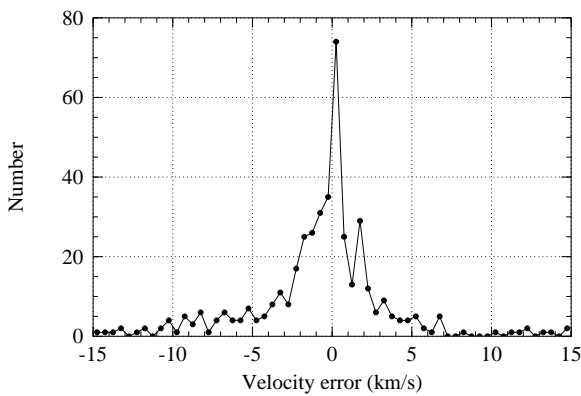}
      \caption{Histogram of radial velocity errors (from run~3) for the model galaxies in bins of $0.5~\mathrm{km \, s}^{-1}$.}
      \label{fig_velocity2}
    \end{center}
  \end{figure}
  
  \subsubsection{Radial Velocity}
  
  The mean velocity error (based on run~3) for the galaxy models is ${-1.9} \pm 54.5~\mathrm{km \, s}^{-1}$. As in the case of source position, the large standard deviation about the mean is caused by galaxies that are only partially detected. By including only sources with position errors of less than $15~\mathrm{arcsec}$ in both right ascension and declination and velocity errors of less than $20~\mathrm{km \, s}^{-1}$ we can exclude such partial detections, resulting in a corrected mean radial velocity error of ${-0.8} \pm 4.6~\mathrm{km \, s}^{-1}$.
  
  A histogram of radial velocity errors for the galaxy models is shown in Figure~\ref{fig_velocity2}. As in the case of point sources, the distribution is not exactly Gaussian. Instead, there is a sharp peak near zero and an underlying broad distribution of errors, in particular in the negative range. Some of these non-Gaussian structures could again be the result of digitisation effects in conjunction with the spectral channel width of $3.86~\mathrm{km \, s}^{-1}$, while we have no conclusive explanation for the noticeable asymmetry of the distribution.
  
  \begin{figure*}[t]
    \begin{center}
      \includegraphics[width=\linewidth]{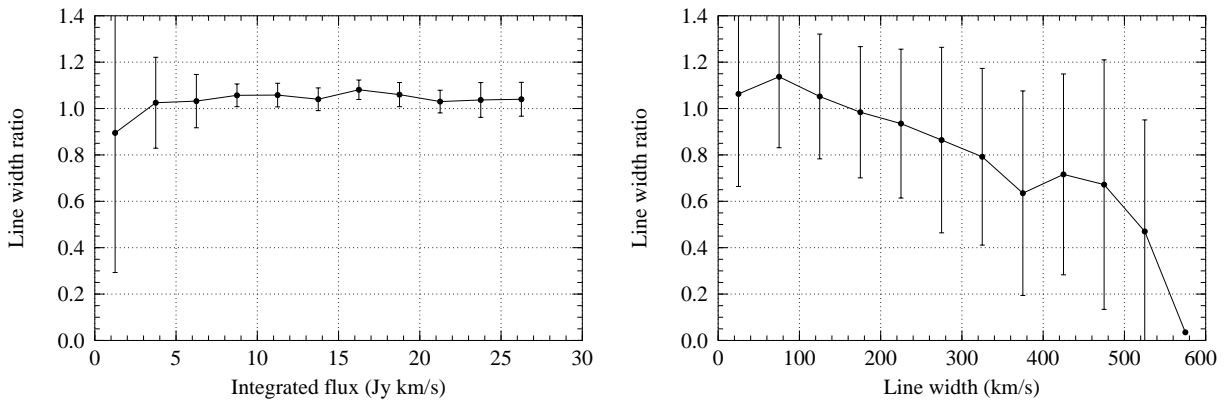}
      \caption{Ratio of measured ($w_{50}$, from run~3) versus true line width for the model galaxies as a function of true integrated flux in bins of $2.5~\mathrm{Jy \, km \, s}^{-1}$ (left-hand panel) and true line width in bins of $50~\mathrm{km \, s}^{-1}$ (right-hand panel). The error bars indicate the standard deviation about the mean.}
      \label{fig_linewidth2}
    \end{center}
  \end{figure*}
  
  \subsubsection{Line Width}
  
  In order to estimate the original line width of the input models, we calculated a `pseudo line width' which balances the intrinsic width of an individual line profile with the overall, integrated line width resulting from the rotation velocity of the galaxy, thus
  \begin{equation}
    w_{\rm mod} = \sqrt{\left[ 2 \, v_{\rm rot} \sin(i) \right]^{2} + w_{\rm int}^{2}}
  \end{equation}
  where $v_{\rm rot}$ is the rotation velocity of the model galaxy, $i$ is the inclination of the disc, and $w_{\rm int} = 22.7~\mathrm{km \, s}^{-1}$ is the intrinsic FWHM of the Gaussian spectral line at each position across the galaxy.
  
  The left-hand panel of Figure~\ref{fig_linewidth2} shows the mean ratio of the measured line width, $w_{50}$, over the calculated `pseudo line width', $w_{\rm mod}$, as a function of true integrated flux in bins of $2.5~\mathrm{Jy \, km \, s}^{-1}$ (based on run~3). \textsc{Duchamp} measures accurate line widths close to the true value over a wide range of fluxes. The small deviation from the value of~$1$ can be easily explained by the fact that $w_{\rm mod}$ is just an approximation to the FWHM of the line profile. Only for fainter sources of $F_{\rm int} \lesssim 5~\mathrm{Jy \, km \, s}^{-1}$ does the line width ratio decrease and the standard deviation increase significantly, indicating larger errors in \textsc{Duchamp}'s measurement of line width.
  
  In the right-hand panel of Figure~\ref{fig_linewidth2} we have plotted the ratio of $w_{50} / w_{\rm mod}$ as a function of $w_{\rm mod}$ in bins of $50~\mathrm{km \, s}^{-1}$. While line width measurements for sources with narrow lines of $w_{\rm mod} \lesssim 250~\mathrm{km \, s}^{-1}$ are on average accurate, there is a systematic discrepancy for sources with broader lines, the line widths measured by \textsc{Duchamp} being systematically too small. The large standard deviation suggests that this could have been caused by cases in which only one half of the galaxy was detected, whereas the other half remained undetected, resulting in a significantly lower value of the measured line width. Nevertheless, line width measurements for fully-detected sources should be accurate even if their line widths are large. This problem again demonstrates the need to identify partially detected sources to avoid systematic errors that would affect the scientific interpretation of the data.
  
  \subsubsection{Integrated Flux}
  
\begin{figure}[t]
    \begin{center}
      \includegraphics[width=\linewidth]{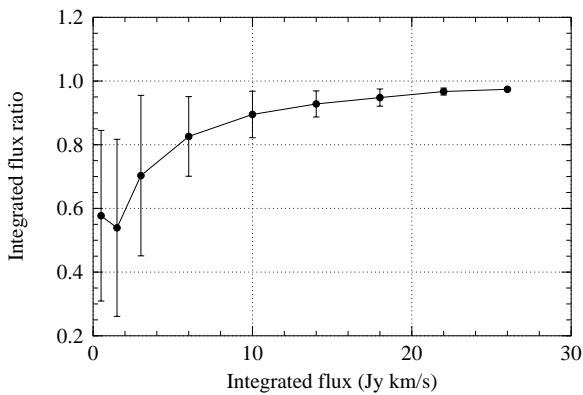}
      \caption{Ratio of measured (from run~3) versus true integrated flux of the model galaxies as a function of true integrated flux. The error bars indicate the standard deviation about the mean. The bin width decreases towards lower fluxes.}
      \label{fig_intflux2}
    \end{center}
  \end{figure}

  The ratio of measured versus true integrated flux of the model galaxies, based on run~3, is shown in Figure~\ref{fig_intflux2}. Similar to our previous tests on point sources (see Figure~\ref{fig_intflux}), the integrated flux measured by \textsc{Duchamp} is systematically too low. For bright sources of $F_{\rm int} \approx 20~\mathrm{Jy \, km \, s}^{-1}$ a large fraction of approximately 95\% of the flux is recovered, whereas this figure drops to below 60\% for fainter sources of $F_{\rm int} \lesssim 2~\mathrm{Jy \, km \, s}^{-1}$. At the same time, the scatter significantly increases, suggesting larger uncertainties (on a relative scale) in the flux measurement of faint sources.
  
  As discussed previously, the reason for the failure of \textsc{Duchamp} to accurately determine the integrated flux of a source is that the software only sums over data elements that are above the flux threshold and hence misses some of the flux. Even the growth of detections down to the $0.5 \sigma$~level has not solved this fundamental problem, although the defect has become less severe than for the point source models without growing (see the right-hand panel of Figure~\ref{fig_intflux} for comparison).
  
  \begin{figure*}[t]
    \begin{center}
      \includegraphics[width=0.9\linewidth]{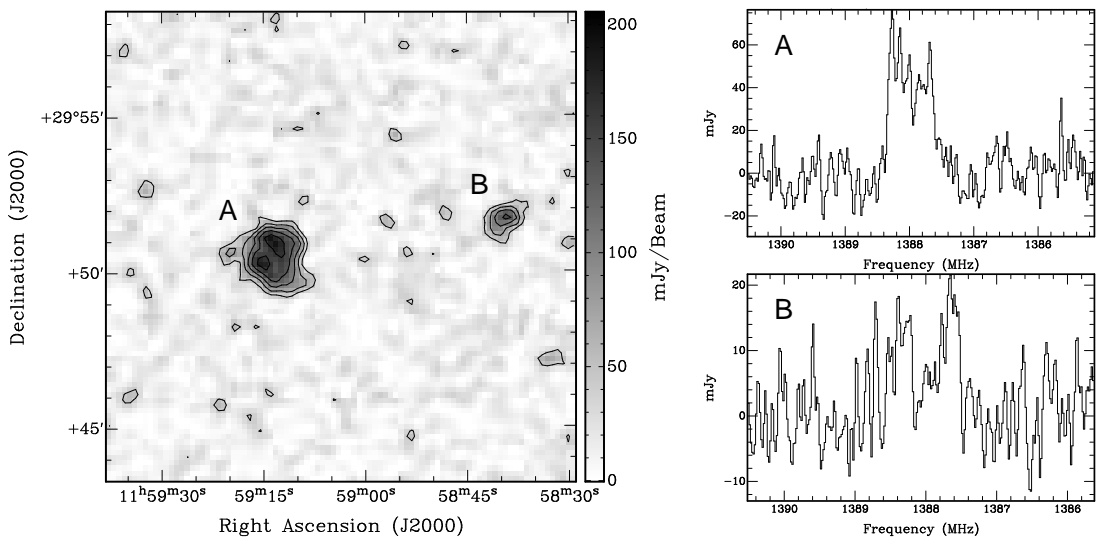}
      \caption{\textit{Left-hand panel:} Moment-zero map of a small region of the WSRT cube with injected WHISP galaxies, showing two galaxies labelled~A and~B. \textit{Right-hand panels:} Integrated spectra of the two galaxies.}
      \label{fig_maps_3}
    \end{center}
  \end{figure*}
  
  \section{Model Cube Based on Real Galaxies}
  \label{sect_whisp}
  
  So far, we have tested \textsc{Duchamp} on artificial sources embedded in perfectly Gaussian noise. While this is useful to study the basic performance of the software, real observations will be more challenging for any source finder due to the more complex morphology of real sources and the presence of various artefacts in the data, e.g.~terrestrial and solar interference, spectral baseline instabilities, or residual continuum emission.
  
  Unfortunately, it is not possible to simply test \textsc{Du\-champ} on a real \ion{H}{i}~data cube, because we would not have a-priori knowledge of the sources in such a cube and would not be able to assess which of the detections made by \textsc{Duchamp} are genuine. A solution to this problem would be to inject copies of real galaxies into a real data cube of ``pure'' noise, i.e.~a data cube extracted from telescopic observations that does not contain any \ion{H}{i}~sources above the noise level. This method combines the advantages of artificial source models, where the source locations and parameters are exactly known, with those of real observations with realistic sources and artefacts.
  
  For this purpose, we generated a data cube containing real noise extracted from an observation with the Westerbork Synthesis Radio Telescope (WSRT). We then added about 100~data cubes from the ``Westerbork Observations of Neutral Hydrogen in Irregular and Spiral Galaxies'' (WHISP) survey \citep{kamphuis1996,swaters2002}, each containing one or more galaxies. The selected WHISP data cube were artificially redshifted by scaling their size and flux level to match sources in a redshift range of $0.02 \lesssim z \lesssim 0.04$, centred on the median redshift of $0.03$ expected for the WALLABY project \citep{koribalski2009}. The procedure for creating the test data cube is explained in more detail by \citet{serra2011}.
  
  The final test data cube has a size of $360 \times 360$~spatial pixels and $1464$~spectral channels. The pixel size of $10~\mathrm{arcsec}$ (with a synthesised beam width of $30~\mathrm{arcsec}$) and channel width of $18.3~\mathrm{kHz}$ (equivalent to about $4~\mathrm{km \, s}^{-1}$) were chosen to reflect the expected specifications of WALLABY. Figure~\ref{fig_maps_3} shows an example image and spectra of two of the galaxies in the final cube. As the locations and properties of the injected galaxies are well-known, we can directly compare them to the output of \textsc{Duchamp} to assess performance indicators such as completeness and reliability.
  
  \subsection{Running {\sc Duchamp}}
  
  \begin{table*}
    \begin{center}
      \caption{Relevant \textsc{Duchamp} input parameters set in the different input parameter files for the model based on real WHISP galaxies and WSRT noise. The default values of \textsc{Duchamp} were used for most of the other parameters. The last two rows list the overall completeness and reliability achieved by \textsc{Duchamp}.}
      \label{tab_input3}
      \begin{tabular}{lrrrrrrrrrrrrrr}
        \hline
        Run               & \hspace{0.3cm} & 1 & 2 & 3 & 4 & 5 & 6 & \hspace{0.3cm} & 7 & 8 & 9 & \hspace{0.3cm} & 10 & 11 \\
	\hline
        threshold         & & $1.0$ & $1.5$ & $2.0$ & $2.5$ & $3.0$ & $3.5$ & & $2.5$ & $3.0$ & $3.5$ & & $2.5$ & $3.0$ \\
        growthThreshold   & &    -- & $1.0$ & $1.0$ & $1.0$ & $1.0$ & $1.0$ & & $1.0$ & $1.0$ & $1.0$ & & $1.0$ & $1.0$ \\
        minPix            & &  $25$ &  $25$ &  $25$ &  $25$ &  $25$ &  $25$ & &  $25$ &  $25$ &  $25$ & &  $25$ &  $25$ \\
        minChannels       & &   $5$ &   $5$ &   $5$ &   $5$ &   $5$ &   $5$ & &   $5$ &   $5$ &   $5$ & &   $5$ &   $5$ \\
        reconDim          & &   $1$ &   $1$ &   $1$ &   $1$ &   $1$ &   $1$ & &   $1$ &   $1$ &   $1$ & &   $3$ &   $3$ \\
        snrRecon          & & $4.0$ & $4.0$ & $4.0$ & $4.0$ & $4.0$ & $4.0$ & & $3.0$ & $3.0$ & $3.0$ & & $4.0$ & $4.0$ \\
        scaleMin          & &   $3$ &   $3$ &   $3$ &   $3$ &   $3$ &   $3$ & &   $3$ &   $3$ &   $3$ & &   $3$ &   $3$ \\
	\hline
	Completeness (\%) & &  $44$ &  $44$ &  $44$ &  $43$ &  $40$ &  $38$ & &  $59$ &  $50$ &  $41$ & &  $49$ &  $41$ \\
        Reliability (\%)  & &  $55$ &  $55$ &  $63$ &  $73$ &  $95$ & $100$ & &  $10$ &  $50$ &  $93$ & &  $12$ &  $88$ \\
        %Break-up (\%)     & &   $2$ &   $2$ &   $2$ &   $2$ &   $3$ &   $0$ & &   $5$ &   $2$ &   $0$ & &   $8$ &   $2$ \\
        \hline
      \end{tabular}
    \end{center}
  \end{table*}
  
  We ran \textsc{Duchamp} multiple times on the WSRT data cube with WHISP galaxies to probe different input parameter settings of \textsc{Duchamp}. A summary of the runs and parameters used is given in Table~\ref{tab_input3}. We mainly covered a wide range of flux thresholds between $1.0$ and $3.5 \sigma$ and tested one-dimensional (spectral domain only) versus three-dimensional (spatial and spectral domain) wavelet reconstruction of the cube. The output catalogue of each run was again cross-matched with the original source catalogue, using the Python script described in Section~\ref{sect_galaxies_duchamp}.
  
  In order to obtain the original source catalogue, we ran \textsc{Duchamp} once on the input model cube without noise, using a very low detection threshold of well below the final noise level and no wavelet reconstruction. This resulted in a list of 100~sources against which the output catalogue provided by \textsc{Duchamp} can be judged. Since this method already introduces a strong bias in the catalogue of source parameters, we will only analyse the completeness and reliability of \textsc{Duchamp}, but we shall not attempt to assess the parametrisation performance of the software, because we do not have an exact source catalogue against which we would be able to assess the source parameters as measured by \textsc{Duchamp} from the final test cube.
  
  \subsection{Results}
  
\begin{figure}[t]
    \begin{center}
      \includegraphics[width=\linewidth]{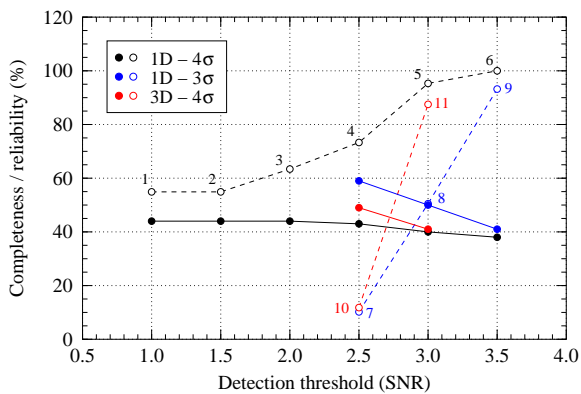}
      \caption{Completeness (filled circles, solid lines) and reliability (open circles, dashed lines) of \textsc{Duchamp} on the WSRT model cube with WHISP galaxies for different flux thresholds and input parameters. The colours, as shown in the legend, distinguish the different wavelet reconstruction modes (one-dimensional versus three-dimensional) and wavelet reconstruction thresholds ($3 \sigma$ versus $4 \sigma$) used in the tests (see Table~\ref{tab_input3} for details). The numbers alongside the data points refer to the corresponding runs as listed in Table~\ref{tab_input3}.}
      \label{fig_whisp}
    \end{center}
  \end{figure}

  Completeness and reliability of the different runs of \textsc{Duchamp} on the WSRT model cube with WHISP galaxies are listed in Table~\ref{tab_input3} and displayed in Figure~\ref{fig_whisp} as a function of detection threshold. Generally, between about $40\%$ to $60\%$ of all galaxies in the cube were found by \textsc{Duchamp}, while the overall reliability varies strongly from about $10\%$ to $100\%$ depending on detection threshold and wavelet reconstruction parameters.
  
  We achieve better results for one-dimensional wave\-let reconstruction (black and blue data points in Figure~\ref{fig_whisp}) which generally yields higher completeness and reliability than three-dimensional wavelet reconstruction (red data points). This is presumably due to the small angular size of most galaxies in the model cube; there is not much to gain from performing a wavelet reconstruction in the spatial domain, whereas one-dimensional wavelet reconstruction in the frequency domain yields much better results because most galaxies are well-resolved and extended in frequency.
  
\begin{figure}[t]
    \begin{center}
      \includegraphics[width=\linewidth]{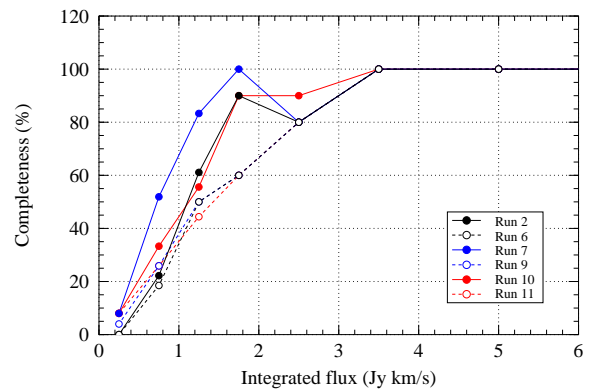}
      \caption{Completeness as a function of integrated flux for selected runs (see legend) of \textsc{Duchamp} on the WSRT model cube with WHISP galaxies. The choice of colours is the same as in Figure~\ref{fig_whisp}.}
      \label{fig_completeness_3}
    \end{center}
  \end{figure}

  In Figure~\ref{fig_completeness_3} we plot completeness as a function of integrated flux for selected runs of \textsc{Duchamp}. Above a flux of $F_{\rm int} \gtrsim 3~\mathrm{Jy \, km \, s}^{-1}$ \textsc{Duchamp} consistently finds all sources irrespective of the input parameters chosen. At lower fluxes the different runs produce significantly different results, with the one-dimensional wavelet reconstruction (black and blue data points) generally performing better than the three-dimensional reconstruction (red data points), as noted before.
  
  The best-performing parameter set in terms of completeness, run~7, produces a completeness of $50\%$ at an integrated flux of $F_{\rm int} \approx 0.7~\mathrm{Jy \, km \, s}^{-1}$, corresponding to an \ion{H}{i}~mass of $1.7 \times 10^{5}~M_{\odot}$ at a distance of $1~\mathrm{Mpc}$, or $1.7 \times 10^{9}~M_{\odot}$ at $100~\mathrm{Mpc}$. This is worse than what we achieved for the point sources with Gaussian line profiles in Section~\ref{sect_pointsources}, but significantly better than the outcome for the model galaxies in Section~\ref{sect_galaxies}. The reason for the better performance could be that the artificially redshifted WHISP galaxies are generally much more compact than the model galaxies created for the tests in Section~\ref{sect_galaxies}. As with any threshold-based source finder, compact sources are easier to detect than extended sources, even with prior wavelet reconstruction or smoothing. As the spectral profiles of the WHISP galaxies are generally broad and complex, performance is worse than in the case of the point source models in Section~\ref{sect_pointsources} which had much simpler and narrower Gaussian lines.
  
  The overall reliability in the case of run~7 is very low with only $10\%$. This figure, however, is the raw reliability achieved by \textsc{Duchamp} and can be substantially improved by filtering sources based on their measured parameters. False detections are usually the result of noise peaks being picked up by the source finder. A large fraction of these false noise detections will be characterised by very low integrated fluxes and small line widths, and often a simple cut in flux--line width space will remove more than $95\%$ of false detections while retaining more than $95\%$ of genuine detections. This fact is illustrated and discussed in more detail in Section~\ref{sect_reliability}.
  
  In summary, when running \textsc{Duchamp} on a realistic data cube with real galaxies at a redshift of about $0.03$ and genuine noise extracted from observational data taken with the WSRT, the software performs as expected with completeness levels ranging in between those achieved for the compact and extended model sources discussed in the previous sections. This result illustrates that the performance of \textsc{Duchamp}, as with any source finder based on flux thresholding, will strongly depend on the morphology and extent of the sources to be detected. Even with multi-scale wavelet reconstruction, \textsc{Duchamp} is more likely to uncover compact sources than sources that are significantly extended, either spatially or spectrally.
  
  At the same time, the performance of \textsc{Duchamp} does not seem to be hampered by the fact that we are dealing with real telescope data and noise, as the completeness and reliability levels reported in Table~\ref{tab_input3} are generally very similar to what we achieved with the model sources discussed in the previous sections. This is presumably due to the excellent quality of the Westerbork data which do not contain any obvious artefacts such as interference or residual continuum emission.

  \section{Discussion}
  \label{sect_discussion}
  
  In general, \textsc{Duchamp} does what it promises to do. It is able to reliably detect sources down to low signal-to-noise ratios and accurately determine their position and radial velocity. These are the most fundamental requirements for any source finder. Our tests also demonstrated that by using and fine-tuning the options of `\`{a} trous' wavelet reconstruction and growing of sources to lower flux levels the performance of \textsc{Duchamp} can be greatly enhanced.
  
  \subsection{Improving Reliability}
  \label{sect_reliability}
  
  The reliability figures reported throughout this paper have all been ``raw'' reliabilities, i.e.~reliabilities as achieved by \textsc{Duchamp} prior to any filtering of the output source catalogue. The user would normally wish to substantially improve these through appropriate filtering of the source catalogue based on the source parameters as measured by \textsc{Duchamp}.
  
\begin{figure*}[t]
    \begin{center}
      \includegraphics[width=\linewidth]{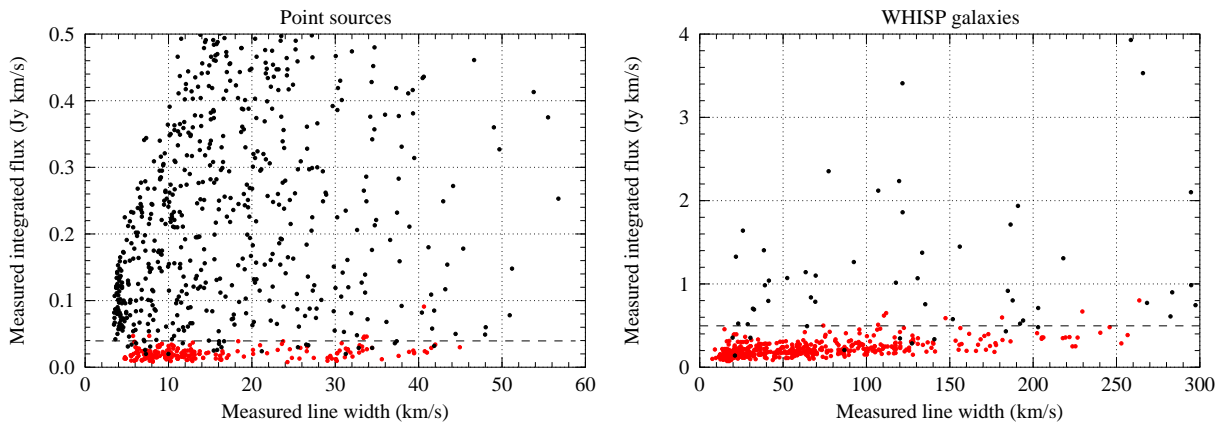}
      \caption{Measured integrated flux, $F_{\rm int}$, versus measured line width, $w_{50}$, of all genuine (black) and false (red) detections made by \textsc{Duchamp} in the point source models with Gaussian line profiles (left) and the test cube with artificially redshifted WHISP galaxies (right). The dashed, black lines indicate the flux levels of $0.04$ and $0.5~\mathrm{Jy \, km \, s}^{-1}$ used to filter false detections.}
      \label{fig_false}
    \end{center}
  \end{figure*}

  The left-hand panel of Figure~\ref{fig_false} shows the measured integrated flux plotted against measured line width for all genuine (black data points) and false (red data points) detections found by \textsc{Duchamp} in the point source models discussed in Section~\ref{sect_pointsources}. It is obvious that genuine and false detections occupy largely disjunct regions of $F_{\rm int}$--$w_{50}$ parameter space, with false detections generally occurring near the low end of the integrated flux spectrum. Similar plots can be generated for other combinations of source parameters, but $F_{\rm int}$ and $w_{50}$ usually provide the best distinction between genuine and false detections.
  
  The easiest way to improve the reliability of \textsc{Du\-champ}'s source finding results is to simply apply a cut in $F_{\rm int}$ to exclude most false detections while retaining most of the genuine sources. In our example, applying a cut at $F_{\rm int} = 40~\mathrm{mJy \, km \, s}^{-1}$ will discard $97.2\%$ of all false detections while at the same time retaining $96.9\%$ of all genuine sources, thereby increasing the overall reliability from $77.1\%$ to $99.2\%$ while only moderately decreasing the overall completeness from $83.0\%$ to $80.6\%$.
  
  A similar cut can be applied to the results from run~7 on the test data cube containing artificially redshifted WHISP galaxies, as plotted in the right-hand panel of Figure~\ref{fig_false}. Again, applying a simple flux threshold of $0.5~\mathrm{Jy \, km \, s}^{-1}$ will improve reliability from $10\%$ to $84\%$, while only moderately decreasing completeness from $59\%$ to $50\%$. The method is not quite as successful as for the point sources, as we are now dealing with real galaxies and real noise with interference and artefacts, but nevertheless a significant improvement in reliability can be achieved without any severe impact on the number of genuine detections.
  
  This simple example illustrates that the ``raw'' reliability figures quoted throughout this paper should not be considered as the final numbers. Reliability can be greatly improved through very basic filtering in parameter space of the \textsc{Duchamp} output catalogue. In principle, this applies to the output of almost any source finder. Alternatively, instead of removing sources from the output catalogue, it may be desirable to calculate a reliability number for each catalogue entry based on the source's location in parameter space and leave it to the catalogue's users to decide as part of their scientific analysis at which reliability level they wish to make the cut.
  
  \subsection{Source Parametrisation Issues}
  
  When it comes to source parametrisation, the measurements provided by \textsc{Duchamp} are affected by several systematic errors. These systematic errors are not due to errors in the software itself, but a consequence of the the presence of noise in the data as well as the methods and algorithms used for measuring source parameters.
  
  Spectral line widths determined by \textsc{Duchamp} are generally very accurate and not much affected by noise-induced, systematic errors as far as the $w_{50}$ parameter is concerned. The two other line width parameters calculated by \textsc{Duchamp}, $w_{20}$ and $w_{\rm vel}$, appear to be systematically too large over a wide range of signal-to-noise ratios and should not be used unless explicitly required in special, well-defined circumstances.
  
  Peak fluxes, as reported by \textsc{Duchamp}, are in general slightly too large for bright sources and significantly too large (on a relative scale) for faint sources. This is due to the fact that \textsc{Duchamp} determines the peak flux by simply selecting the value of the brightest pixel encountered. This method introduces a bias towards positive noise peaks sitting on top of the brightest region of a source, and hence, in the presence of noise, peak fluxes measured by \textsc{Duchamp} will be systematically too high.
  
  Integrated fluxes determined by \textsc{Duchamp} are significantly and systematically too small, in particular for faint sources. This is likely caused by the fact that \textsc{Duchamp} simply sums over the flux of discrete elements above a given threshold to determine the integrated flux, thereby missing some of the flux from elements below the flux threshold. Hence, the raw integrated flux measurements currently provided by \textsc{Duchamp} are not useful and need to be corrected to compensate for the systematic offset. This issue is particularly sensitive as many scientific projects, including the ASKAP survey science projects WALLABY and DINGO\footnote{\textit{Deep Investigation of Neutral Gas Origins}; principal investigator: Martin Meyer; public website: http:/\slash{}internal.physics.uwa.edu.au\slash{}$\sim$mmeyer\slash{}dingo/} \citep{meyer2009}, rely on accurate flux measurements, for example for determining the \ion{H}{i}~mass function of galaxies.
  
  Finally, a particular problem in the case of galaxies is that under certain circumstances galaxies either get broken up into multiple detections or only one half of a galaxy is detected. This problem mainly affects faint, edge-on galaxies with broad spectral profiles that are partly hidden in the noise and results in systematic errors in the measurements of essentially all source parameters, including basic parameters such as position and radial velocity. Such cases of multiple or partial detections must be identified and treated separately to prevent biases in any scientific analysis based on the source finding results.
  
  \section{Summary}
  \label{sect_summary}
  
  In this paper we present and discuss the results of basic, three-dimensional source finding tests with \textsc{Du\-champ}, the standard source finder for the Australian SKA Pathfinder, using different sets of unresolved and extended \ion{H}{i}~model sources as well as a data set of real galaxies and noise obtained from \ion{H}{i}~observations with the WSRT.
  
  Overall, \textsc{Duchamp} appears to be a successful, gen\-eral-purpose source finder capable of reliably detecting sources down to low signal-to-noise ratios and accurately determining their position and velocity. In the case of point sources with simple Gaussian spectral lines we achieve a completeness of about 50\% at a peak signal-to-noise ratio of~3 and an integrated flux level of about $0.1~\mathrm{Jy \, km \, s}^{-1}$. The latter corresponds to an \ion{H}{i} mass sensitivity of about $2 \times 10^{8}~\mathrm{M}_{\odot}$ at a distance of $100~\mathrm{Mpc}$ which is slightly better than what the WALLABY project is expected to achieve for real galaxies \citep{koribalski2009}. The situation is less ideal for extended sources with double-horn profiles. In this case we achieve 50\% completeness at an integrated flux level of about $2.5~\mathrm{Jy \, km \, s}^{-1}$ for the model galaxies and $0.7~\mathrm{Jy \, km \, s}^{-1}$ for the WHISP galaxies. The latter is equivalent to an \ion{H}{i} mass sensitivity of about $1.7 \times 10^{9}~\mathrm{M}_{\odot}$ at a distance of $100~\mathrm{Mpc}$, illustrating that the performance of \textsc{Duchamp}, as well as any other source finder, will strongly depend on source morphology. However, these figures may well be improved by carefully optimising the various input parameters offered by \textsc{Duchamp}.
  
  In its current state \textsc{Duchamp} is not particularly successful in parametrising sources in the presence of noise in the data cube, and other, external algorithms for source parametrisation should be considered instead. It appears, however, that most, if not all, para\-metrisation issues are due to intrinsic limitations in the implemented algorithms themselves and not due to errors in their implementation, suggesting that most of the problems can in principle be solved by implementing more sophisticated parametrisation algorithms in \textsc{Duchamp}. Alternatively, corrections would have to be applied to all parameters derived by \textsc{Duchamp} to compensate for systematic errors. Such corrections, however, would have to be highly specialised and tailored to the particular survey and source type concerned.
  
  \section*{Acknowledgments}
  
  We wish to thank Matthew Whiting for creating the \textsc{Duchamp} source finder and for having had an open ear for our comments and suggestions regarding improvement of the software.

  %\end{multicols}
  
\end{document}